\documentclass[aps,prd,preprint,groupedaddress]{revtex4-1}
\usepackage{graphicx}
\usepackage{epstopdf}
\usepackage{amsmath,amssymb}
\usepackage{bm}

\begin{document}

\vspace{0mm}


\title{Flavor structure of the unpolarized and longitudinally-polarized
sea-quark distributions in the nucleon}


\author{M.~Wakamatsu}
\email[]{wakamatu@phys.sci.osaka-u.ac.jp}
\affiliation{Department of Physics, Faculty of Science, \\
Osaka University, \\
Toyonaka, Osaka 560-0043, Japan}



\begin{abstract}
It is now widely recognized that a key to unravel the
nonperturbative chiral-dynamics of QCD hidden in the deep-inelastic-scattering
observables is the flavor structure of sea-quark distributions in the nucleon.
We analyze the flavor structure of the nucleon sea
in both of the unpolarized and longitudinally polarized parton distribution functions (PDFs) within a single theoretical framework
of the flavor SU(3) chiral quark soliton model (CQSM),
which contains only one adjustable parameter $\Delta m_s$,
the effective mass difference between the strange and nonstrange
quarks. A particular attention is paid to a nontrivial correlation between
the flavor asymmetry of the unpolarized and longitudinally polarized sea-quark
distributions and also to a possible particle-antiparticle asymmetry
of the strange quark distributions in the nucleon.
We also investigate the charge-symmetry-violation (CSV) effects in the parton
distribution functions exactly within the same theoretical framework, which
is expected to provide us with valuable information on the relative importance
of the asymmetry of the strange and antistrange distributions and the CSV
effects in the valence-quark distributions inside the nucleon in the
resolution scenario of the so-called NuTeV anomaly in the extraction of
the Weinberg angle.
\end{abstract}


\maketitle


\section{Introduction}

The nucleon structure function physics works out in a fine
balance of the perturbative and nonperturbative quantum chromodynamics (QCD).
The standard approach to the deep-inelastic-scattering (DIS) physics is
based on the so-called factorization theorem, which states that the
DIS scattering cross section is factorized
into two parts, i.e. the hard part which is tractable within the
framework of perturbative QCD and the soft part containing the information of
nonperturbative nucleon structure \cite{Muta98}\nocite{Collins11}
\nocite{AEL95}-\cite{LR00} . Customarily, the soft part is treated
as a black box, which should be determined through experiments.
This is certainly a reasonable strategy.
We however believe that, even if this part is completely fixed by experiments, 
one would still want to know why those PDFs take the forms so determined.
Furthermore, we now realize that a key ingredient to reveal the
nonperturbative chiral dynamics of QCD is hidden in the soft part of the
DIS physics in the form of the flavor structure (or flavor dependence)
of the sea-quark (or anti-quark) distributions \cite{NMC91}
\nocite{Kumano98}\nocite{GP01}\nocite{PQ14}\nocite{Sullivan72}\nocite{Aubert83}
\nocite{Kumano91}\nocite{KL91}\nocite{Wakamatsu91}\nocite{Wakamatsu92}
\nocite{HSB91}-\cite{KFS96}.
Unfortunately, what can be extracted from the well-founded inclusive DIS
analyses are only the combinations of quark plus sea-quark (or anti-quark)
distributions.
To separate out anti-quark distributions, we need either of neutrino-induced
DIS scattering measurements, semi-inclusive DIS (SIDIS) measurements
or Drell-Yan measurements. Because of the smallness of the neutrino-induced
DIS cross section, here we are forced to use nuclear targets, which inevitably
introduces large theoretical uncertainties in addition to statistical errors
arising from the smallness of the event counting rate of neutrino-induced
reactions \cite{CCFR95}\nocite{CCFRNuTeV01}\nocite{BPZ00}\nocite{NNPDF09}
\nocite{AKP09}\nocite{SYKMOO08}-\cite{SYKKMOO09}.
On the other hand, a lot of efforts have been made for understanding the
SIDIS mechanism \cite{HERMES08}\nocite{HERMES05}\nocite{HERMES99}
\nocite{COMPASS09}-\cite{COMPASS08}, in particular, the fragmentation
mechanism of a quark or an antiquark into observed
hadrons \cite{Kretzer00}\nocite{DSSV07A}\nocite{DSSV07B}\nocite{HKNS07}
\nocite{AKK08}-\cite{Belle08}.
Still, one must say that
our understanding of the the semi-inclusive reaction mechanism remains
at fairly lower level than that of inclusive reactions.     
A complementary approach to DIS physics is necessary here to clarify
possibly important role of chiral dynamics of QCD in the DIS physics,
based on effective models of QCD or on lattice QCD.

Although there are lots of models of baryons, the chiral quark soliton model (CQSM), first proposed by Diakonov, Petrov and Pobylitsa, would probably be the
best one \cite{DPP88}, at least as an effective model of internal partonic
structure of the baryons including the nucleon.
(The practical numerical method for
handling the CQSM was established in \cite{WY91} based on the general
methodology of Kahana and Ripka \cite{KR84}. The unique feature of the
CQSM, which plays an important role in the so-called nucleon spin problem,
was also pointed out in this paper. For early reviews of the CQSM, see
\cite{Review_Wakamatsu92}\nocite{Review_CBKPWMAG96}
\nocite{Review_ARW96}-\cite{Review_DP01}.)
The CQSM has a lot of merits over other effective models of baryons.
First, it is a relativistic mean-field theory of quarks, with inclusion of
infinitely many Dirac-sea orbitals, which means that it is a field
theoretical model including infinitely many dynamical degrees of freedom.
Second, the mean-field is of hedgehog-shape in harmony with the nonperturbative dynamics expected from large $N_c$ QCD.
One interesting consequence of this unique feature of the
model is strong spin-isospin correlation (or anti-correlation) in the
generated nucleon seas \cite{WY91}.
Third, in association with the first advantage, its field theoretical nature
enables us reasonable estimation not only
of quark distributions but also of antiquark distributions
\cite{DPPPW96}\nocite{DPPPW97}\nocite{PPGWW99}\nocite{WGR96}\nocite{WGR97}
\nocite{GRW98}\nocite{WK98}\nocite{WK99}\nocite{WW00}\nocite{Wakamatsu03A}-
\cite{Wakamatsu03B}.
Last but not least, only one parameter of
the flavor SU(2) version of the model. i.e the dynamically generated quark
mass $M$, is already fixed to be $M \simeq 375 \,\mbox{MeV}$ from low
energy phenomenology as well as from theoretical
ground \cite{DPP88}. To handle
the strange-quark degrees of freedom in the nucleon, we must extend the model
to flavor SU(3). However, this flavor SU(3) extension of the model need to
introduce only one additional parameter, i.e. the mass difference between
the strange and nonstrange quarks \cite{Wakamatsu03A},\cite{Wakamatsu03B}.
This means that we can still make nearly parameter-free predictions for 
parton distribution functions (PDFs).
This should be contrasted with variant species of meson cloud (convolution)
models, which are also believed to incorporate the nonperturbative chiral
dynamics of QCD.
In fact, the meson cloud models contain quite a few
model parameters such as several meson-quark coupling
constants, coupling form factors, parameters of parton distributions in the
mesons, and so forth \cite{CCS10},\cite{CS03}. 
Moreover, the model predictions often depend critically on how many
meson-baryon intermediate states are included in the theoretical
calculations. This last fact is sometimes a serious obstruction for giving
unique and quantitatively trustable predictions on the sea-quark
distributions in the nucleon.
We emphasize again that the CQSM does not suffer from these bothersome
problems, because it is nearly parameter-free.
As a matter of course, the biggest problem or shortcoming common in all
the low energy effective models of the nucleon including the CQSM is a lack
of explicit gluon degrees of freedom. This point should always be kept in
mind when applying low energy models of the nucleon to the DIS physics,
as we shall discuss later.

The main purpose of the present paper is to unravel the nonperturbative
chiral dynamics of QCD hidden in the parton distribution functions of the
nucleon through the analysis of the flavor structure of the nucleon seas.
An important point is that the flavor structure of the nucleon seas including
the strange quark degrees of freedom is analyzed simultaneously for the
unpolarized PDFs and for the longitudinally polarized PDFs within
a single theoretical framework.
It is the flavor SU(3) version of the CQSM, which contains only
one adjustable parameter, i.e. the effective mass difference
between the strange and nonstrange quarks.
To get a feeling about the reliability of the model, we first carry out
a systematic comparison between the model predictions and the results of
the most recent unbiased global fits by the NNPDF group.
In our opinion, this systematic comparison is of special importance.
This is because, if one picks up only a specific distribution functions,
it would not be extremely difficult to reproduce the corresponding
empirical distributions, especially by the models like the meson cloud
models containing many parameters as well as freedoms.
Unfortunately, it sometimes happens that such an agreement
is fortuitous and the same model with the same set of parameters fails
to reproduce other independent distributions.
This is the reason why we believe it important to check how well
a particular model can or cannot reproduce wide class of empirical PDFs
simultaneously.  

We also investigate the charge-symmetry-violating (CSV) effects in the parton
distribution functions based on exactly the same theoretical framework.
The motivation to investigate the CSV effects
in the nucleon parton distributions is as follows. It is known that, to resolve
the widely-known anomaly on the Weinberg angle of the electroweak standard
model raised in the analysis of the neutrino-induced DIS measurements
by the NuTeV group \cite{NuTeV02A},\cite{NuTeV02B}, the two mechanisms
of QCD origin are believed to play important roles.
They are the asymmetry of the strange and anti-strange quark distributions
in the nucleon and the CSV effects in the valence-quark distributions in the
proton and the neutron.
Which of these ingredients is more important is not a completely settled
issue \cite{ST87}\nocite{BW92}\nocite{BM96}\nocite{DM04}\nocite{AI04}
\nocite{Wakamatsu05}\nocite{LT05}\nocite{LPT10}\nocite{Sather92}
\nocite{RTL94}\nocite{LT03}\nocite{GJR05}\nocite{MRST04}
\nocite{MST06}-\cite{CSSM11}.
We believe that the analysis of the $s$-$\bar{s}$ asymmetry and the CSV
distributions within a single theoretical framework would provide us with
a valuable information on the relative importance of these two mechanisms.  

The paper is organized as follows.
First, in sect.II, through the comparison of the predictions of the SU(3)
CQSM for the unpolarized PDFs with the recent global fits by the
NNPDF group \cite{NNPDFNLO2.1},
we try to estimate the degrees of reliability of the model.
After that, we concentrate on inspecting the characteristic
feature of the model predictions for flavor structure of the unpolarized light-flavor sea quark distributions, which has not been
determined very reliably on the observational basis alone.
The characteristic predictions of the SU(3) CQSM will be compared with
the predictions of other models of the sea-quark distributions in the
nucleon as well as with other empirical information if available.
In sect.III, a similar analysis is carried out
for the longitudinally polarized PDFs, the global analyses of which
was recently reported by NNDPD group \cite{NNPDFpol1.0}.
Next, in sect.IV,
we investigate the CSV effects in the light-flavor quark and anti-quark
distributions within exactly the same framework of the SU(3) CQSM.
A main emphasis there is put on getting useful information on the
relative importance of the CSV effects and the strange and anti-strange
quark asymmetry in the resolution of the NuTeV anomaly for the Weinberg
angle. Then, we summarize what we have learned on the
flavor structure of the nucleon seas in sect.V.  


\section{flavor SU(3) CQSM and unpolarized PDFs}

The theoretical formulation of the flavor SU(3) CQSM for evaluating
the PDFs in a baryon was already described
in detail in our previous papers \cite{Wakamatsu03A},\cite{Wakamatsu03B}.
(The SU(3) CQSM itself was first proposed in \cite{BDGPPP93}.)
We therefore give here only a brief
sketch of it by confining to its basic theoretical structure. 
The model lagrangian of the flavor SU(3) CQSM is a straightforward
extension of the SU(2) one. It is given by
\begin{equation}
 {\cal L} \ = \ {\cal L}_0 \ + \ {\cal L}_{SB},
\end{equation}
where
\begin{equation}
 {\cal L}_0 \ = \ \bar{\psi} (x) \,\left( \,i \! \not\!\partial \ - \
 M \,U^{\gamma_5} (x) \right) \,\psi (x) ,
\end{equation}
with
\begin{equation}
 U^{\gamma_5} (x) \ = \ e^{\,i \,\gamma_5 \,\pi (x) / f_\pi} ,
 \hspace{8mm} \pi (x) \ = \ \pi_a (x) \,\lambda_a \ \
 \ (a = 1, \cdots, 8) ,
\end{equation}
being the SU(3) symmetric part of the lagrangian, while
\begin{equation}
 {\cal L}_{SB} \ = \ - \,\,\bar{\psi} (x) \,\Delta m_s \,P_s \,\psi (x) ,
\end{equation}
with
\begin{eqnarray}
 \Delta m_s \,P_s \ = \
 \left( \begin{array}{ccc}
 \ 0 \ & \ 0 \ & \ 0 \  \\
 \ 0 \ & \ 0 \ & \ 0 \  \\
 \ 0 \ & \ 0 \  &\  \Delta m_s  
 \end{array} \right) ,
\end{eqnarray}
being the SU(3) breaking part resulting from the mass
difference between the strange and non-strange quarks. (Here, we neglect
the light-quark masses so that $\Delta m_s \equiv m_s - m_{u,d} = m_s$.) 
The above effective lagrangian contains eight meson fields instead of three
in the case of the flavor SU(2) model.
However,  we recall that, in the framework of the CQSM, these meson
fields are not independent fields of quarks,
as inferred from the fact that there is no kinetic term of the meson
fields in the above basic lagrangian of the model.

Now, fundamental dynamical assumptions of the model is as follows.

\begin{itemize}

\item First, the lowest energy classical solution (or the mean-field solution)
is obtained by the embedding of SU(2) mean-field solution of hedgehog
shape into the SU(3) matrix.
(The same dynamical assumption is also used in more familiar 
SU(3) Skyrme model \cite{Witten83}\nocite{NMP84}-\cite{Guadagnini85}.)

\item Second is the SU(3) symmetric quantization of the rotational motion
in the collective coordinate space. 

\item The third is the perturbative treatment of the SU(3) symmetry breaking
mass term. We recall that this mass difference $\Delta m_s$
between the strange and nonstrange quarks is {\it only one parameter} of
the model.

\end{itemize}

We fix this only one parameter of the model is as follows.
It is taken as an adjustable parameter between the physically reasonable
range $m_s = (80 - 120) \,\mbox{MeV}$. As a general rule,
the distribution functions for the light-flavor $u$- and $d$-quarks are
generally rather insensitive to the value of $\Delta m_s$.
As naturally expected,
what are most sensitive to the value of
$\Delta m_s$ is the strange-quark distributions. 
We found that overall good reproduction of the shape of the empirical
strange quark distribution $s(x) + \bar{s}(x)$ is obtained with the
choice $\Delta m_s = 80 \,\mbox{MeV}$. We therefore fix the value
of $\Delta m_s$ to be $80 \,\mbox{MeV}$ and continue to use it throughout all
the following calculations. This means that there remains no more free
parameter in the model.   

Before going to the discussion on the predictions of the SU(3) CQSM for
the unpolarized parton distribution functions (PDFs) in comparison with
the empirical information given at the high energy scales, we think it
important to explain our general strategy for applying an effective
model to DIS physics.
It is widely believed that the predictions of effective models of hadrons
should be taken as those given in the low energy domain of nonperturbative
QCD, while the parton distribution functions extracted from
experiments correspond to high energy scale of perturbative QCD.  
A difficult question is how to harmonize two domains of QCD.
It is customarily assumed that the model predictions for PDFs given at the
low energy scale can be related through QCD evolution equation to empirically
extracted PDFs at high energy. 
The central difficulty we encounter here is a matching scale problem.
That is, it is far from trivial how to specify the exact model energy scale
from which one starts the evolution as above.
Most effective models of baryons like the MIT bag model or the meson
cloud models use fairly low starting energy $Q^2_{ini} \simeq 0.16 \,\mbox{GeV}^2$.
On the other hand, there is some argument that the starting energy of
the CQSM should be taken to be a little higher.
In fact, we recall here the argument by Petrov et al. based on the instanton
picture of the QCD vacuum \cite{DP86},\cite{Diakonov03},
which is thought to give a theoretical foundation of the CQSM.
According to them, the scale of the CQSM is set
by the inverse of the average instanton size $\rho$ as
$Q_{ini} \sim 1 \,/\,\rho \sim 600 \,\mbox{MeV}$.
Although reasonable, it seems to us that the relation between the
choice of initial scale and the average instanton size is very qualitative.
It just indicates that any choice between $Q^2_{ini} \simeq 0.3 \,\mbox{GeV}^2$
and $Q^2_{ini} \simeq 0.4 \,\mbox{GeV}^2$ would be equally well.
A fully satisfactory choice of the initial energy scale of evolution
would be obtained only when one carries out a proper renormalization
procedure of nonperturbative QCD, as is actually done in the framework
of lattice QCD, although the calculation of the PDFs is not yet possible
in this promising framework. (Another advantage of the lattice QCD treatment
is that the renormalization is carried out at fairly high-energy scale,
i.e. $Q^2 = 4 \,\mbox{GeV}^2$, where one can safely starts the perturbative
evolution to higher energy scales.)
Even though there is a theoretical indication that the model scale of the
CQSM is higher than those of other effective models of baryons, it is
still much smaller than the scale of $1 \,\mbox{GeV}$, so that some
sensitivity of the final predictions on the choice of the initial scale
of evolution cannot be completely avoided.
For instance, we find that the two choices
$Q^2_{ini} = 0.30 \,\mbox{GeV}^2$ and $Q^2_{ini} = 0.40 \,\mbox{GeV}^2$
cause difference in the range of $(4 \sim 8) \,\%$ for the heights of
valence-like peaks of the unpolarized PDFs at $Q^2 = 2 \,\mbox{GeV}^2$.
Since better agreement with the empirical PDFs is obtained with the
choice $Q^2_{ini} = 0.30 \,\mbox{GeV}^2$, we continue to use this value,
which was the value used in our previous studies \cite{WK99}\nocite{WW00}
\nocite{Wakamatsu03A}-\cite{Wakamatsu03B}.

\begin{figure}[ht]
\begin{center}
\includegraphics[width=8.5cm]{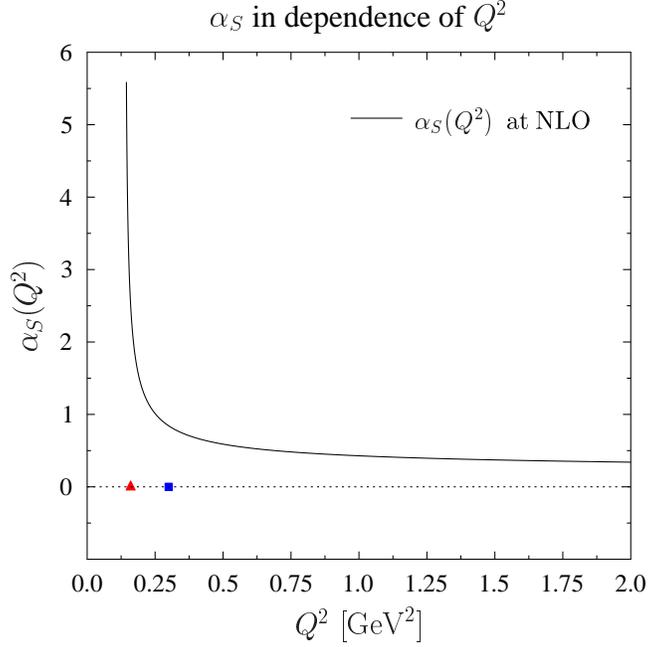}
\caption{The QCD running coupling constant
$\alpha_S \equiv g^2 \,/\,4 \,\pi$ at the NLO
in dependence of $Q^2$. The filled triangle (red in color) corresponds to
the frequently used starting energy scale of evolution in the MIT bag model
or meson cloud models, whereas the filled square (blue in color) to the
starting energy scale used in the CQSM.}
\label{Fig01_alphas_NLO}
\end{center} 
\end{figure} 

In any case, we emphasize again that the value
$Q^2_{ini} = 0.30 \,\mbox{GeV}^2$ which we use as the initial scale
of evolution in the CQSM is a little higher than the value
$Q^2_{ini} \simeq 0.16 \,\mbox{GeV}^2$ frequently used in many effective
models of baryons like the MIT bag model or the meson cloud models. 
This difference is sometimes critical, because the validity of using
perturbative evolution equation at too low energy scales is a delicate
question. In fact, we show in Fig.\ref{Fig01_alphas_NLO}
the QCD running coupling constant at the next-to-leading order (NLO) as a
function of $Q^2$. 
(Here we have used the exact solution of the NLO evolution equation with
the standard $\overline{\rm MS}$ scheme in the fixed-flavor scheme with
$n_f = 3$ even beyond the charm threshold. However, the effects of charm
on the quantity discussed here would be very small as compared with the
necessary precision of our discussion.)
One sees that the $\alpha_S$ at the scale
of $Q^2 = 0.16 \,\mbox{GeV}^2 = (400 \,\mbox{MeV})^2$ already shows
a diverging behavior, which throws a little doubt on the use of the
perturbative renormalization group equation at such scales.
On the other hand, at the initial energy scale of the CQSM, i.e. at
$Q^2 = 0.30 \,\mbox{GeV}^2 \simeq (550 \,\mbox{MeV})^2$, the
perturbative QCD may be barely applicable. 
(Whether the value of $\alpha_S \simeq 0.84$ at the scale
$Q^2 = 0.30 \,\mbox{GeV}^2$ is large or small is a delicate question.
However, more transparent measure of the applicability of the
perturbative renormalization group equation is provided by the change
rate of $\alpha_S$ as a function of $Q^2$, which can be easily read in. 
Another remark is that, if one uses the LO evolution
equation, the diverging behavior of $\alpha_S (Q^2)$ appears at
lower energy scale. This is one of the reason why many low energy models
like the MIT bag model or the meson cloud models adopt the LO evolution
equation together with very low starting energy of evolution.
However, since the main purpose of our analysis is
to compare the predictions of the SU(3) CQSM with the NNPDF fits
carried out at the NLO, the consistency requires to use the evolution
scheme at the NLO.)

\begin{figure}[ht]
\begin{center}
\includegraphics[width=9.0cm]{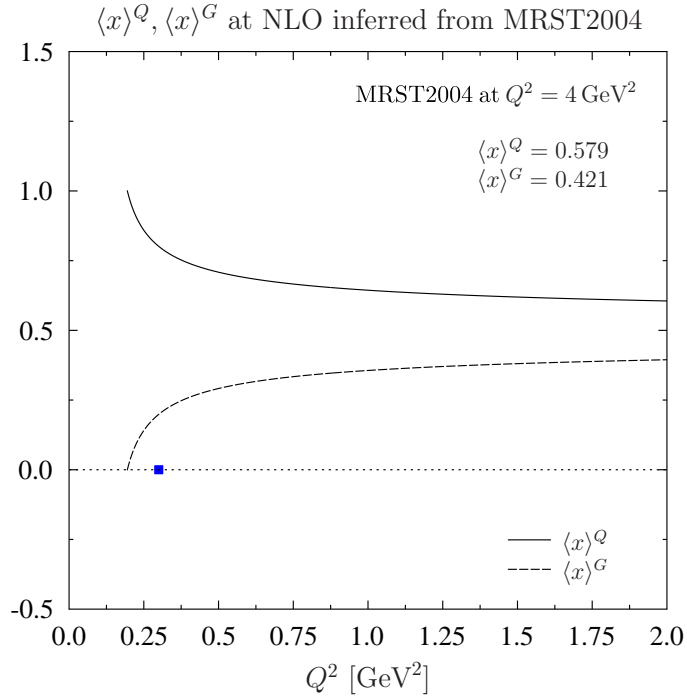}
\caption{The quark and gluon longitudinal momentum fractions as functions
of $Q^2$, obtained by solving the QCD evolution equation at the NLO with
the initial conditions $\langle x \rangle^Q = 0.579$ and
$\langle x \rangle^G = 0.421$ given at $Q^2 = 4 \,\mbox{GeV}^2$ by the
MRST2004 analysis \cite{MRST04},\cite{MST06}.
The filled square (blue in color) corresponds to the starting energy scale
used in the CQSM.}
\label{Fig02_xQ_xg_NLO}
\end{center} 
\end{figure}

As shown above, although our choice of a little higher starting energy
of evolution is preferable from the standpoint of using the scheme of
perturbative renormalization group equation, there is one thing which
we must pay attention to. The key quantities in our argument here
are the momentum fractions of quarks and gluons as functions of
the energy scale $Q^2$. Up to this time, the momentum
fractions of quarks and gluons in the nucleon at the high-energy scale are
fairly precisely known.
Given below are the empirical values for the quark and gluon momentum
fractions $\langle x \rangle^Q$ and $\langle x \rangle^G$ given
at $Q^2 = 4 \,\mbox{GeV}^2$ by the MRST2004
analysis \cite{MRST04},\cite{MST06} :
\begin{equation}
 \langle x \rangle^Q = 0.579, \ \ \ \langle x \rangle^G = 0.421 . 
\end{equation}
Here, for simplicity, we have neglected very small error bars.
As an interesting trial, we carried out a {\it downward evolution} of the quark
and gluon momentum fraction, by starting with these known empirical values
at the high energy scale. The results are respectively shown by the solid and
the long-dashed curves in Fig.\ref{Fig02_xQ_xg_NLO}.

\vspace{4mm}

As anticipated, as $Q^2$ decreases, the quark momentum fraction
$\langle x \rangle^Q$ increases, whereas the gluon momentum fraction
$\langle x \rangle^G$ decreases to eventually become zero at a certain
energy scale. An important observation here is the fact that, at
the model energy scale of the CQSM, i.e.
$Q^2 = 0.30 \,\mbox{GeV}^2 \simeq (550 \,\mbox{MeV})^2$, the gluon
still carries about 20 \% of the nucleon momentum.
Since the CQSM is an effective quark model, which does not contain
explicit gluon degrees of freedom, to start the evolution at
$Q^2 = 0.30 \,\mbox{GeV}^2$ amounts to neglecting important role
of gluons, which are likely to carry about 20 \% of the nucleon
momentum even at this relatively low energy scale.
We will show later that this observation has an important
phenomenological consequence in the interpretation of the
predictions of the CQSM evolved to the high energy scales. 

Before proceeding further, it would be fair to refer to another
limitation of the CQSM. The limitation is due to general restriction
from the limit of large numbers of colors $N_c$. As argued by
Diakonov et al. \cite{DPPPW96},\cite{Diakonov03}, the CQSM provides a
practical realization of large-$N_c$ QCD, so that the parton distribution
functions depend on the Bjorken variable $x$ in such a way that
$N_c \, x = O(1)$ in the limit $N_c \rightarrow \infty$.
Since $N_c = 3$ in nature, this dictates that the CQSM is a good
approximation to QCD only in the region of ``not too small'' $x$
in order to comply with the above scaling law. More concrete
argument on the applicability range of $x$ was given in the paper
by Petrov et al. \cite{PPPBGW98}. Since the CQSM is an effective theory
of QCD, which is not renormalizable, it needs a physical cutoff.
An effective regularization energy $\Lambda_{cut}$ is provided by
the inverse of the average instanton size $\rho$ as
$\Lambda_{cut} \sim 1 \,/\, \rho \sim 600 \,\mbox{MeV}$.
Petrov et al. argued that, in the region
$x \leq (M \,/\,\Lambda_{cut}) \,/\,N_c \simeq 0.1$, the model predictions
for the parton distributions are sensitive to the cutoff energy and/or
the detail of the regularization method, so that they are not necessarily
reliable. In the following study, we take less ambitious pragmatic
standpoint that the CQSM is one of the effective models of baryons
like the MIT bag model or the meson cloud models, and will show the
predicted PDFs in the whole range of $x$, i.e. $0 < x < 1$, although
the above caution should be kept in mind.

Now we are in a position to compare the predictions of the SU(3) CQSM
for the unpolarized PDFs with the empirically extracted ones.
First, to get a feeling about the degrees of success or failure of the model,
we compare our predictions with the recent unbiased global fits
of unpolarized PDFs by the NNPDF group. (Here, we use the NNPDF NLO2.1
fits at the NLO with $n_f = 3$ \cite{NNPDFNLO2.1}.)
The NNPDF fits are given at $Q^2 = 2 \,\mbox{GeV}^2$ for the following
combinations of the PDFs : 

\begin{itemize}

\item the singlet distribution, $\Sigma (x) \equiv \sum_{i=1}^{n_f} \,
(q_i (x) + \bar{q}_i (x))$,
\item the gluon, $g(x)$,

\item the total valence, $V (x) \equiv \sum_{i=1}^{n_f} \,
(q_i (x) - \bar{q}_i (x))$,

\item the nonsinglet triplet, $T_3 (x) \equiv (u(x) + \bar{u}(x)) - 
(d(x) + \bar{d}(x))$,

\item the sea asymmetry distribution, $\Delta_S (x) \equiv \bar{d}(x) - 
\bar{u}(x)$,

\item the strange anti-strange sum, $S^+ (x) \equiv s(x) + \bar{s}(x)$,

\item the strange anti-strange difference, $S^- (x) \equiv s(x) - \bar{s}(x)$.

\end{itemize}

For making a comparison, the CQSM predictions given at the initial scale
$Q^2_{ini} = 0.30 \,\mbox{GeV}^2$ are evolved to the corresponding scale
of $Q^2 = 2 \,\mbox{GeV}^2$ by using the evolution equations at
the next-to-leading order (NLO). 

\vspace{4mm}
\begin{figure}[ht]
\begin{center}
\includegraphics[width=15.0cm]{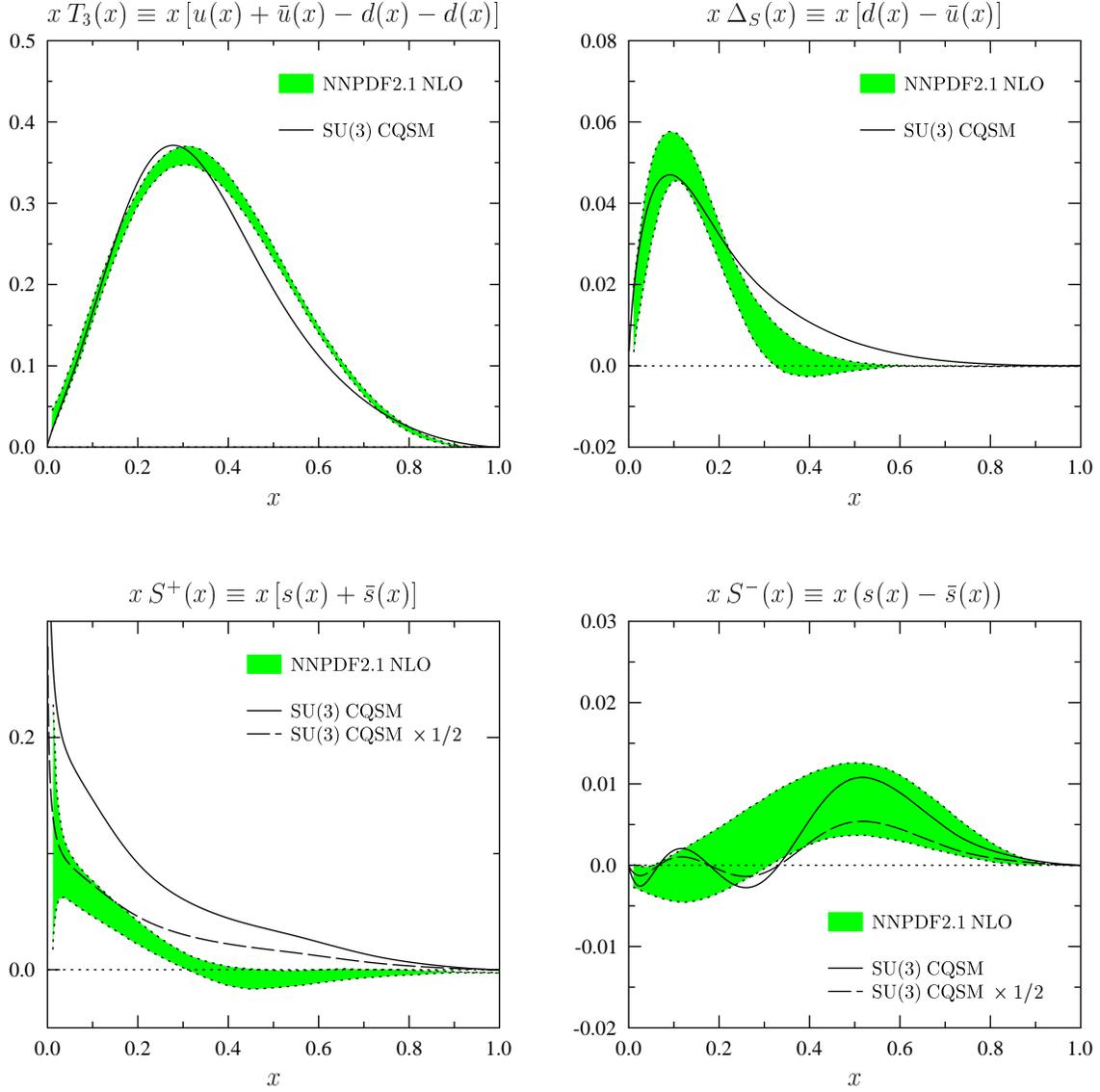}
\caption{The predictions of the SU(3) CQSM for the nonsiglet and strange
quark distributions evolved from $Q^2_{ini} = 0.30 \,\mbox{GeV}^2$ to
$Q^2 = 2 \,\mbox{GeV}^2$ in comparison with the NNPDF2.1 NLO global fits
shown by the shaded areas \cite{NNPDFpol1.0}. The solid
curves are the predictions of the SU(3) CQSM. The long-dashed curves
for the strange quark distribution are the reduced predictions of the
SU(3) CQSM by a factor of $1/2$ explained in the text.}
\label{Fig03_Allunpol_nsinglet}
\end{center} 
\end{figure}

Fig.\ref{Fig03_Allunpol_nsinglet} show the comparison for the PDFs,
$x \,T_3 (x)$, $x \,\Delta_S (x)$,
$x S^+ (x)$, and $x \,S^- (x)$.
We find fairly good agreement between the theory and the NNPDF fits
for the flavor nonsinglet triplet distribution $T_3 (x)$ and the light-flavor
sea asymmetry $\Delta_S (x)$. The detailed inspection reveals that 
the agreements are not perfect. However, in view of almost parameter-free
nature of the model, this agreement can be
taken as one of the nontrivial successes of the CQSM, which properly
takes account of the chiral dynamics of QCD.
(Incidentally, we stress that these distributions $T_3 (x)$ and $\Delta_S (x)$
are quite insensitive to the value of $\Delta m_s$.)

Turning to the strange distributions, we find that the model prediction for
$S^+ (x) = s(x) \,+ \,\bar{s}(x)$ appears to overestimate the NNPDF fit roughly
by a factor of two. This feature of the model prediction could be anticipated.
As was intensively discussed for the SU(3) Skyrme model, the SU(3)
symmetric collective quantization supplemented with the perturbative
treatment of the SU(3) breaking mass difference term, which is taken over to
our treatment of the SU(3) CQSM, has a danger of overestimating the effects
of kaon clouds, which might in turn lead to an overestimation of the strange
quark components in the nucleon \cite{YA88}\nocite{CHK88}-\cite{BRS88}.
We conjecture that plausible predictions for the strange and anti-strange
distributions in the nucleon would lie just between the predictions of the
SU(3) CQSM and the SU(2) CQSM, which amounts to multiplying a
reduction factor $1/2$ to the SU(3) CQSM predictions for $s(x)$
and $\bar{s}(x)$.
As a matter of fact, the reduction factor of just $1/2$ has no strict
foundation and rather ad hoc.
It can be any number between $1$ and
$0$. In principle, this reduction factor can be treated as additional
parameter of the model. However, there is no absolutely trustworthy
empirical information to fix this parameter. We therefore simply
say that, as seen from Fig.\ref{Fig03_Allunpol_nsinglet},
after multiplying this reduction factor of $1/2$, the model prediction
for $S^+ (x)$ is order of magnitude consistent with the current NNPDF fit,
except for larger $x$ region, where the
NNPDF fit does not necessarily respect the positivity of the
distribution. 

Also very interesting is the asymmetry
of strange and anti-strange distributions. 
Noteworthy feature of the NNPDF fit is that the difference distribution
$x \,S^- (x) \equiv x \,[s(x) - \bar{s}(x)]$ has a peak around $x \sim 0.5$.
Very curiously, this feature is perfectly consistent with the prediction
of the SU(3) CQSM. The good agreement is not limited to the position
of the peak. The absolute magnitude of the asymmetry is also consistent
with the NNPDF fit. Note that the bare prediction
of the SU(3) CQSM and the reduced prediction by a factor of $1/2$ are
both consistent with the NNPDF fit within the uncertainty band, although
we prefer the reduced prediction.

\begin{figure}[t]
\begin{center}
\includegraphics[width=12.5cm,height=17.5cm]{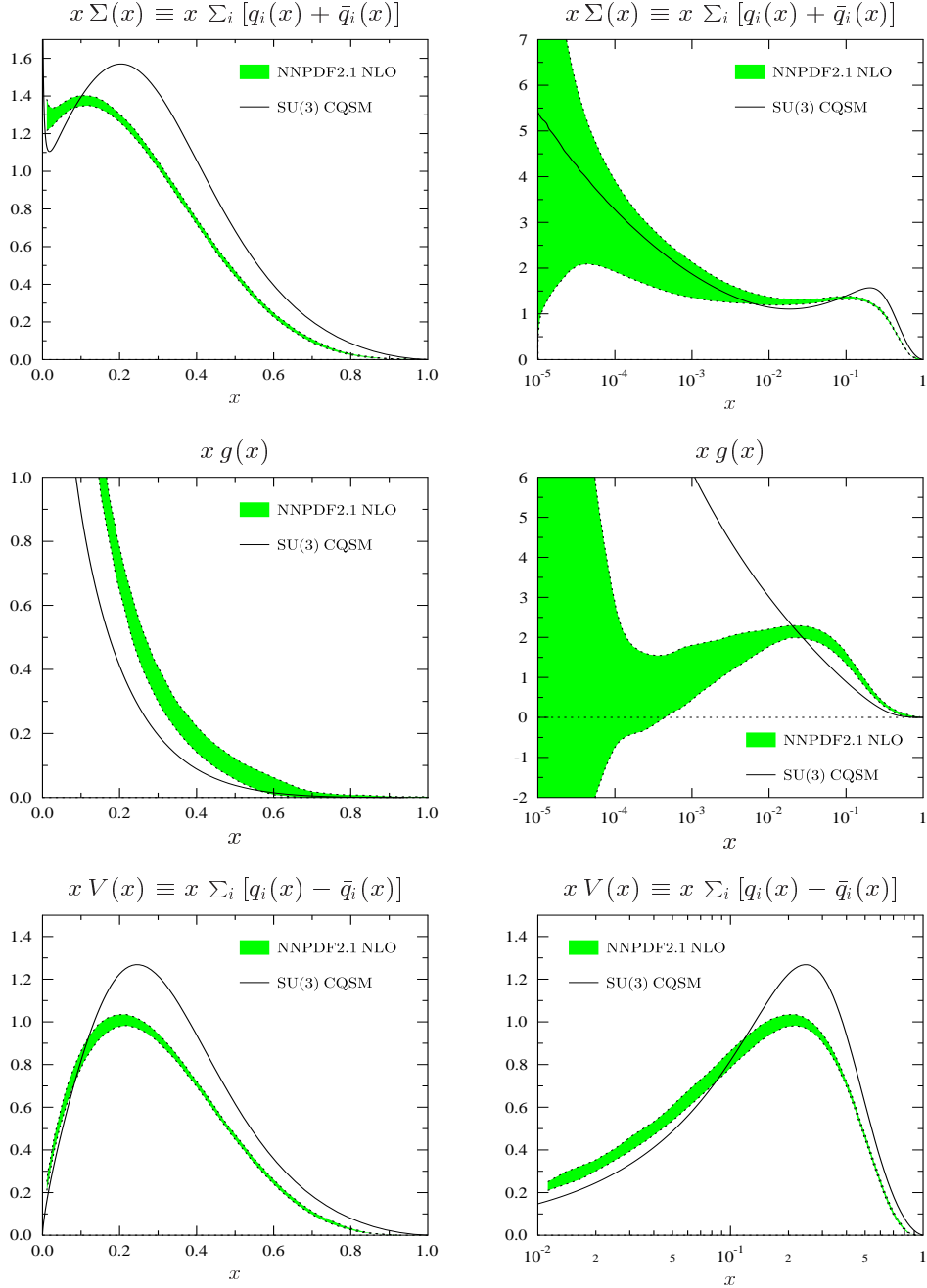}
\caption{The SU(3) CQSM predictions for the singlet quark, gluon and valence
quark distributions in comparison with the NNPDF2.1 NLO fits.}
\label{Fig04_Allunpol_singlet}
\end{center} 
\end{figure} 

Next, in Fig.\ref{Fig04_Allunpol_singlet},
we show the model predictions for the singlet distribution
$\Sigma (x)$, the gluon distribution $g(x)$, and the net valence
distribution $V(x)$, in comparison with the NNPDF fits.
As compared with the success for the nonsinglet distributions, we find
that the model prediction overestimates the NNPDF fit by about 20 \%.
The reason of this discrepancy may be interpreted as follows.
We already pointed out that, at the starting energy scale of evolution,
the gluon field is likely to carry about 20 \% of the total nucleon
momentum, which means that the quark fields carry only about
80 \% of the nucleon momentum. On the other hand, the CQSM is an effective
quark model, which does not contain explicit gluon degrees of freedom,
the net nucleon momentum is naturally saturated by the momenta of
quarks and anti-quarks at the model scale.
Thus, we simply had to set the gluon distribution to be zero
at the starting energy scale of evolution, i.e. at
$Q^2_{ini} = 0.30 \,\mbox{GeV}^2$. This naturally fails to take
account of the fact that the net momentum fraction of quarks at
the initial scale must be only about 80 \%, which would then lead to
an overestimation of the flavor singlet combination of the
quark and anti-quark distribution $\Sigma (x)$ by about 20 \%.    
By the same reason, we cannot expect that the model can give a
reasonable description of the gluon distribution even though nonzero
gluon distribution is generated through evolution. Naturally,
the gluon distribution obtained in this way has no valence-like peak
as observed in the empirical fit.

Turning to the total valence distribution $V(x)$, one again observes
that the model prediction overestimates the NNPDF fit by about
20 \%. The reason of this overestimation is
slightly more complicated than the singlet distribution $\Sigma (x)$.
Since this distribution $V(x)$ is given as a difference of the quark and
anti-quark distributions, it does not couple to the gluon distribution
at the process of scale evolution different from the distribution
$\Sigma (x)$.    
Still, since it is a symmetric sum of the three flavors, $u$, $d$, and $s$,
the possible overestimation of the net distribution at the initial
energy scale pointed out before is likely to remain also at higher
energy scales. Another possible reason would be
that the model might still underestimate the sum of the light-flavor
sea quark distributions, i.e., $\bar{u}(x) + \bar{d}(x)$, which leads
to the overestimation of the combination
$u (x) - \bar{u}(x) + d(x) - \bar{d}(x) + s(x) - \bar{s}(x)$, provided that the
contribution of $s(x) - \bar{s}(x)$ in this combination is small. 
    
A lesson learned from the above analysis is as follows. The
overall agreement between the SU(3) CQSM and the NNPDF2.1 NLO fits
are fairly good in view of nearly parameter-free nature of
the model predictions. However, the agreement is not naturally perfect.
The main reason of discrepancy would be the neglect of the gluon degrees
of freedom, which appears to play non-negligible roles in the flavor-singlet
channel even at relatively low energy scales.
On the other hand, we shall see in the next section that the role of gluons
at the low energy model scale of the CQSM is likely to be much less important
in the case of longitudinally polarized distributions.  

\begin{figure}[ht]
\begin{center}
\includegraphics[width=8.0cm]{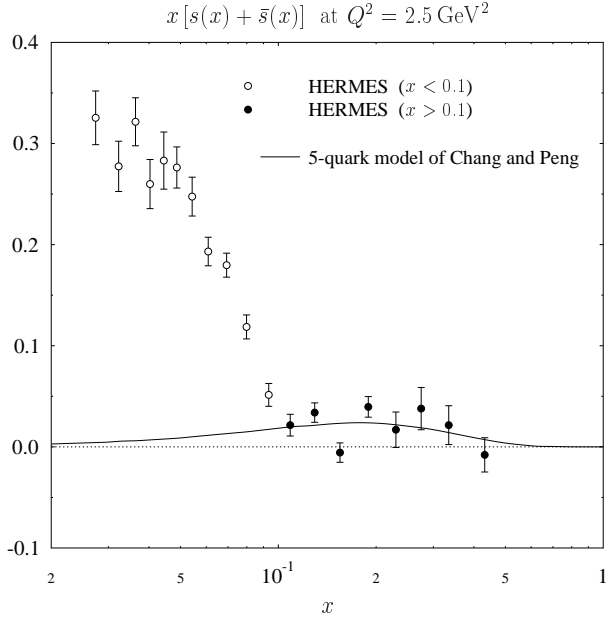}
\caption{HERMES strange quark distribution \cite{HERMES08} in comparison with
the prediction of the 5-quark model of Chang and Peng \cite{CP11A},
evolved to $Q^2 = 2.5 \,\mbox{GeV}^2$ from the initial scale of
the model $Q^2_{ini} = 0.25 \,\mbox{GeV}^2$.}
\label{Fig05_xs_Chang_Peng}
\end{center} 
\end{figure}

After having got a feeling on the degrees of reliability of the model
as well as its limitation, through the comparison with the unbiased global
fits of the unpolarized PDFs by the NNPDF groups,
we now turn our attention to more detailed inspection
on the flavor structure of the sea-quark (anti-quark) distributions
in the nucleon. To unravel the underlying physics, particularly
instructive here is a comparison with related theoretical investigations
as well as other experimental information if available. 
First, we call attention to the strange distribution $s(x) + \bar{s}(x)$
in the nucleon extracted from the analysis of charged kaon production in
semi-inclusive DIS (SIDIS) by the HERMES group \cite{HERMES08}.
As is widely known,
the extracted $s(x) + \bar{s}(x)$ distribution appears to have
intriguing two-component structure as illustrated
in Fig.\ref{Fig05_xs_Chang_Peng}.
Here, following the paper \cite{CP11A}, the HERMES data with $x < 0.1$ are
represented by the open circles,
while those with $x > 0.1$ are by the filled black circles just by
the reason of guidance for eye.

\vspace{4mm}

The observed two-peaked structure motivates Chang and Peng to introduce
an interesting physical interpretation to be explained below \cite{CP11A}. 
(See also more recent paper by Chang, Cheng, Peng and Liu \cite{LCCP12}.) 
Their interpretation is based on the idea of intrinsic charm in the
nucleon proposed many years ago by Brodsky, Hoyer, Petesen, and
Sakai \cite{BHPS80},\cite{BPS81} (BHPS model).
According to BHPS, the intrinsic sea is a component
that is expected to have valence-like peak at larger $x$, while the
extrinsic sea is thought to be generated through QCD splitting processes.
Inspired by this idea, Chang and
Peng et al. proposed an idea that $x > 1$ HERMES data are dominated by
``intrinsic'' sea, while $x < 0.1$ data from ``extrinsic'' sea \cite{CP11A}.
According to them, a component of the HERMES data, which
has a peak around $x \sim (0.1 - 0.3)$ can be reproduced by the
intrinsic 5-quark model (see the solid curve in Fig.\ref{Fig05_xs_Chang_Peng})
with the mixing
rate $P^{uuds\bar{s}}_5 \simeq 0.024$ of the 5-quark component in the
nucleon. At first sight, this appears to provide a reasonable explanation of
the peak structure of $x \,[s(x) + \bar{s}(x)]$ in the higher $x$ region.
However, the following
question immediately arises. Admitting that the account of the 5-quark
component nicely explains the peak structure at higher $x$, how can one
explain the sea-like component in the lower $x$ domain? What is
important to recognize here is the fact that the solid curve
in Fig.\ref{Fig05_xs_Chang_Peng} shows the theoretical prediction,
which was obtained after taking account of the evolution effects by solving
the evolution equation
starting from the scale $Q^2_{ini} = 0.25 \,\mbox{GeV}$.
This means that, to explain the whole HERMES data including the
lower $x$ behavior, one absolutely need
a significant sea-like component already at the starting energy scale.
What generate these sea-like components ?
Assuming the correctness of the HERMES extraction, they must be
higher Fock-components of the nucleon state like the 7-quark component,
5-quark plus gluon, and so on. It is not absolutely clear whether the
valence-like peak structure of the strange quark distribution, which is
obtained by confining to the lowest 5-quark Fock-component only, will
remain or not after taking account of all the these higher Fock-components
of the nucleon wave function.

\vspace{4mm}
\begin{figure}[ht]
\begin{center}
\includegraphics[width=14.0cm]{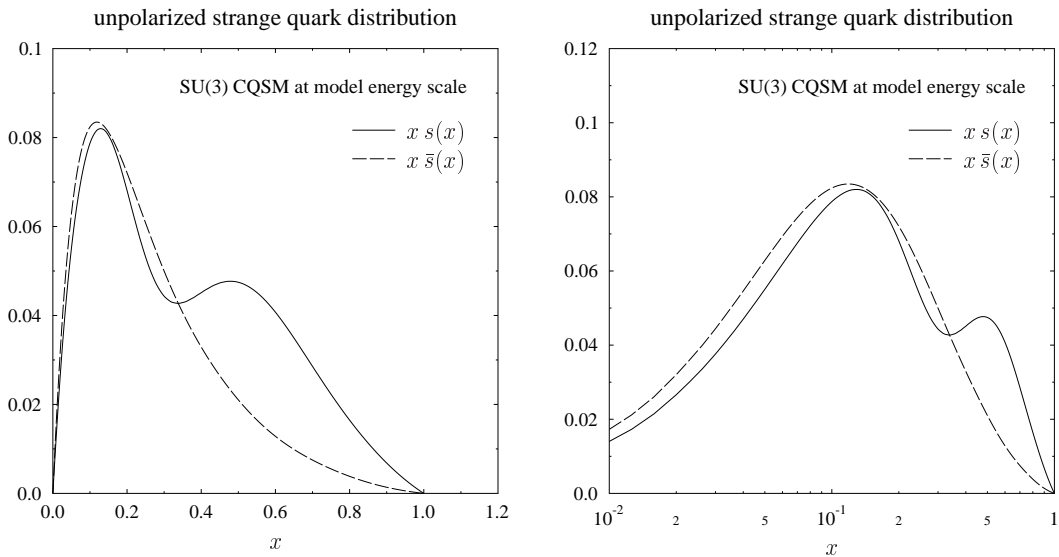}
\caption{The SU(3) CQSM predictions for the strange and anti-strange
distribution in the nucleon at the model energy scale. The left panel
is in linear scale in $x$, while the right panel is in log scale.}
\label{Fig06_ssb_CQSM}
\end{center} 
\end{figure}

To answer the question raised above, we find it useful
to look into the prediction of the SU(3) CQSM for the strange
and anti-strange distribution functions at the low energy model scale.
They are shown in Fig.\ref{Fig06_ssb_CQSM}.
Very interestingly, the model predicts a sizable difference between
the strange and anti-strange quark distributions. 
The strange quark
distribution $x \,s(x)$ shows a two-component structure, i.e. the valence-like
peak in the higher $x$ region and the sea-like component in the lower $x$ region.
On the other hand, the anti-strange quark distribution has only
a sea-like peak in the lower $x$ region. In particular,
one finds that  the $s$-quark distribution has larger $x$
component than the $\bar{s}$-quark distribution. Very interestingly,
this is just a feature
expected from the kaon cloud model of the nucleon advocated
by Signal and Thomas \cite{ST87}, Burkardt and Warr \cite{BW92}, and
also by Brodsky and Ma \cite{BM96} many years ago.
According to the kaon cloud picture, the strange and anti-strange quark
distributions in the proton are generated through the virtual dissociation process
$p \rightarrow \Lambda + K^+$. In this virtual intermediate states, 
the $s$-quark is contained in a baryon, i.e. in $\Lambda$, while
the $\bar{s}$-quark is contained in a meson, i,e. in $K^+$.
This is expected to explain that the $s$-quark has
valence-like harder component than $\bar{s}$-quarks \cite{BW92},\cite{BM96}.
Although we believe that this meson cloud picture gets straight to the point
in a qualitative sense, its quantitative predictability cannot be
expected too much, because the meson cloud models generally contain
too many adjustable parameters and large ambiguities.
A great advantage of the CQSM is that it does not assume any
explicit meson-baryon intermediate states like the nucleon and pion, the
nucleon and rho meson, and the lambda and kaon, etc. Note that 
the above difference between the strange and anti-strange quark
distributions is an automatic consequence of almost parameter-free
calculation. Here, it is very important to recognize the fact that not only
the valence-like component but also the sea-like component are
generated as a consequence of solving the bound state
equation of the nucleon. In this sense, one can call the latter too
as ``intrinsic'' sea not ``extrinsic'' sea, even though
it is not a component with valence-like character. 
The point is that the basic theoretical framework of
the CQSM is the mean-field theory (followed by the collective
quantization of the zero-energy rotational modes), which enables us to
incorporate infinitely many higher multi-quark
components in the language of perturbative Fock-space expansion.
This argument indicates that the decomposition of the quark seas
into the ``intrinsic'' and the ``extrinsic'' components
is a strongly model-dependent or theoretical-scheme-dependent idea.

\vspace{3mm}
\begin{figure}[ht]
\begin{center}
\includegraphics[width=14.0cm]{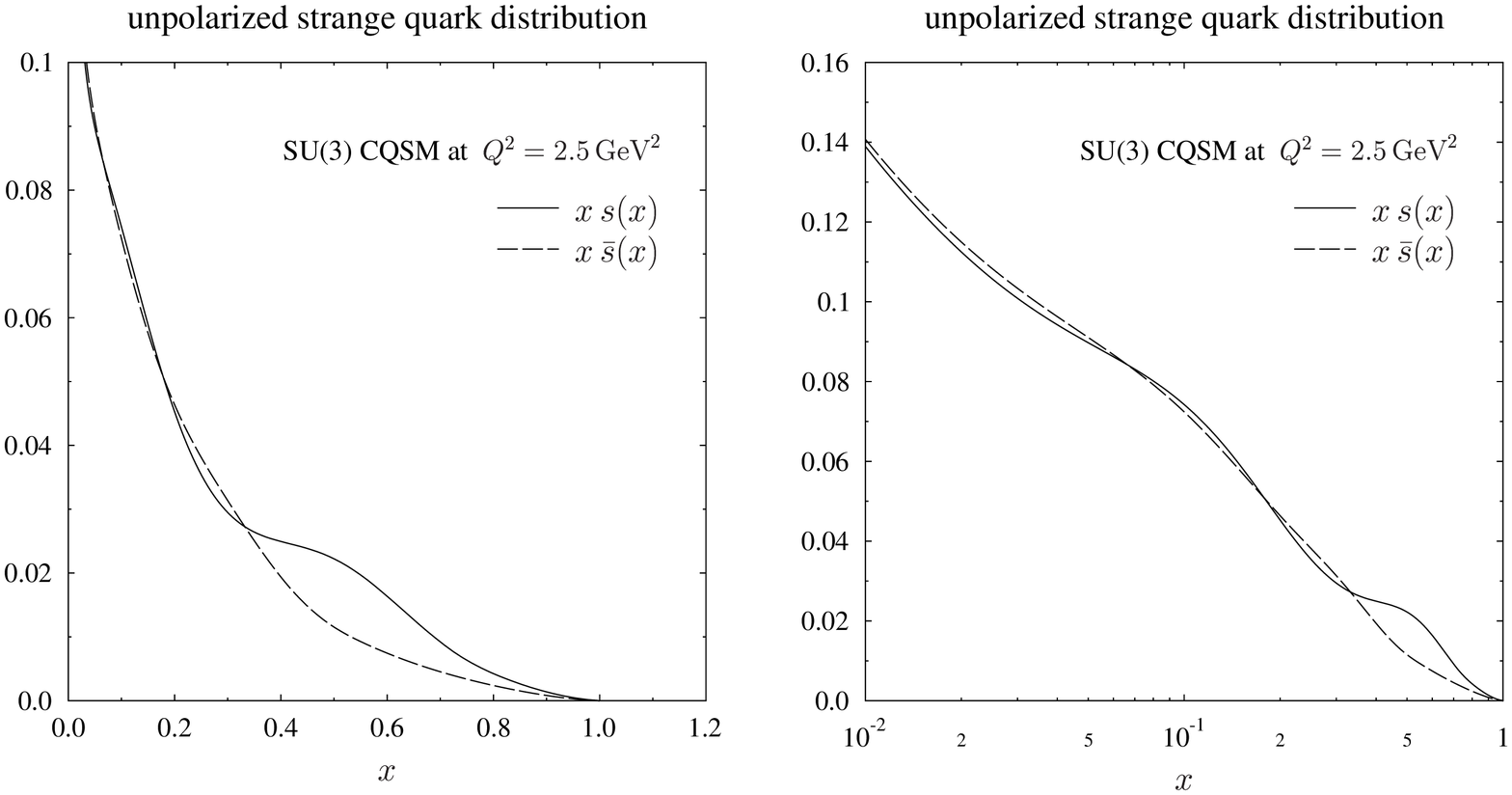}
\caption{The SU(3) CQSM predictions for the strange and anti-strange
distribution in the nucleon evolved to $Q^2 = 2.5 \,\mbox{GeV}^2$.
The left panel is in linear scale in $x$, while the right panel is
in log scale.}
\label{Fig07_ssb_CQSM_evol}
\end{center} 
\end{figure}

Also interesting here is the effect of evolution. We show in
Fig.\ref{Fig07_ssb_CQSM_evol} the prediction of the SU(3) CQSM
evolved to $Q^2 = 2.5 \,\mbox{GeV}^2$
corresponding to the HERMES SIDIS extraction of the distribution
$x \,[s(x) + \bar{s}(x)]$.
One sees that the trace of the valence-like peaked structure of the
distribution $x \,s(x)$ still remains faintly.
However, it is smoothly connected to the sea-like structure in the
lower $x$ domain. Accordingly, we do not see clear two-component
structure in $x \,s(x)$ any more.
On the other hand, since the distribution $x \,\bar{s}(x)$
has only a sea-like component even at the low energy model scale,
the evolved distribution is simply sea-like.
  
\begin{figure}[ht]
\begin{center}
\includegraphics[width=10.0cm]{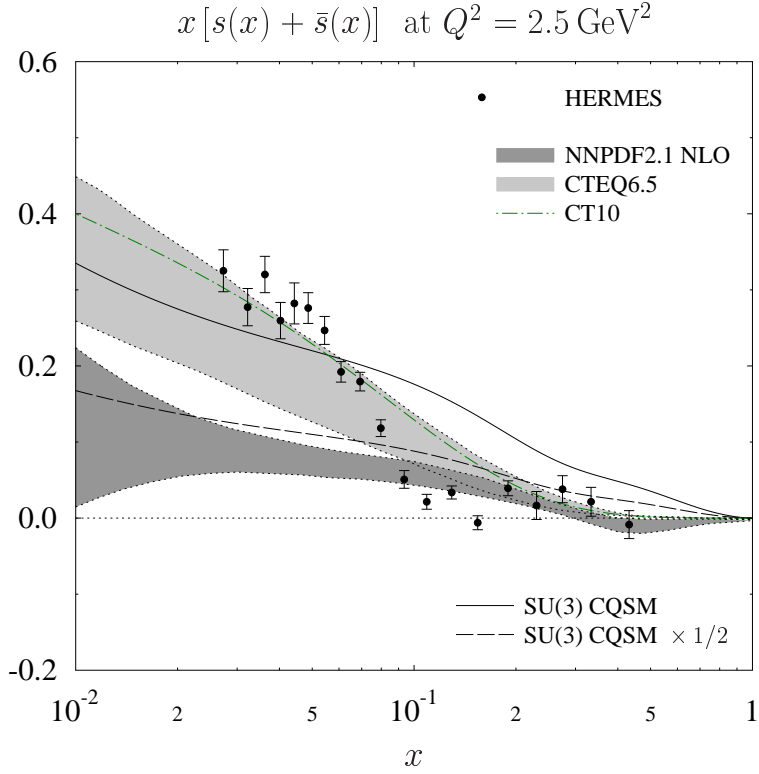}
\caption{The SU(3) CQSM predictions for the strange and anti-strange
distribution at $Q^2 = 2.5 \,\mbox{GeV}^2$ in comparison with
the HERMES SIDIS extraction \cite{HERMES08}.
The solid curve stands for the bare prediction
of the SU(3) CQSM, whereas the dashed curve is the reduced prediction of
the same model by a factor of $1/2$.
The global fits by the NNPDF group given at $Q^2 = 2 \,\mbox{GeV}^2$ and
the two global fits by the CTEQ group corresponding to
$Q^2 = 2.5 \,\mbox{GeV}^2$ are also shown for
comparison \cite{CTEQ08},\cite{CT10}.}
\label{Fig08_xs_Sep_HERMES}
\end{center} 
\end{figure} 

After these considerations, it is instructive to
compare the prediction of the strange plus anti-strange quark
distribution with the corresponding HERMES extraction as well as
several other global fits. The filled black circles in
Fig.\ref{Fig08_xs_Sep_HERMES} stand for the
HERMES SIDIS extraction for $x \,[s(x) + \bar{s}(x)]$.
The thicker shaded area represents the NNPDF global fit given at
$Q^2 = 2.0 \,\mbox{GeV}^2$, while the thinner shaded area does the
CTEQ6.5 fit corresponding to $Q^2 = 2.5 \,\mbox{GeV}^2$.
 (Here, the NNPDF fit corresponds to slightly lower
scale, but the effect of this difference is expected to be small as compared
with sizably large difference with the CTEQ6.5 fit.)
The newer CT10 fit is also shown for reference by the dash-dotted curve
for reference.
The bare prediction of the SU(3) CQSM is shown by the solid curve,
whereas the reduced prediction of the SU(3) CQSM is by the long-dashed
curve. (By the reason already explained, we prefer the reduced
prediction for the strange quark distributions.)
As pointed out above, the SU(3) CQSM prediction
for $x \,[s(x) + \bar{s}(x)]$ at $Q^2 = 2.5 \,\mbox{GeV}^2$
does not show any clear two-component structure, which is indicated
by the HERMES data. Note that this is also a common feature of all the
global fits including the NNPDF group and the CTEQ group.
As is well known, the
HERMES extraction of the strange distribution heavily depends on
the expectation that our understanding of the
semi-inclusive charged-kaon production mechanism is robust enough.
Actually, the small-$x$ data in the HERMES extraction corresponds
to relatively low energy kinematical region, say, $Q^2 \sim 1 \,\mbox{GeV}^2$,
where one would generally expect fairly large higher twist corrections
to the DIS analysis. Still another problem pointed out by Leader, Sidorov,
and Stamenov is that the HERMES analysis uses the factorized QCD
treatment of the data in kinematical region where it is not necessarily
justified \cite{LSS14}. We also point out that the most recent HERMES
analysis \cite{HERMES14}, which is claimed to confirm their earlier
analysis \cite{HERMES08}, was criticized in a recent paper by
Stolarski \cite{Stolarski14}.
Stolarski emphasized the importance of carrying out a careful analysis
in which not only the multiplicity sum of the kaon but also that of the pions
as well as other combinations of $K^+$ and $K^-$ multiplicities are
analysed simultaneously. In any case, we strongly feel that some totally
independent extraction of the strange quark distributions, for example, by
using the neutrino-induced inclusive DIS measurements, is highly desirable. 

\vspace{4mm}
\begin{figure}[ht]
\begin{center}
\includegraphics[width=16.0cm]{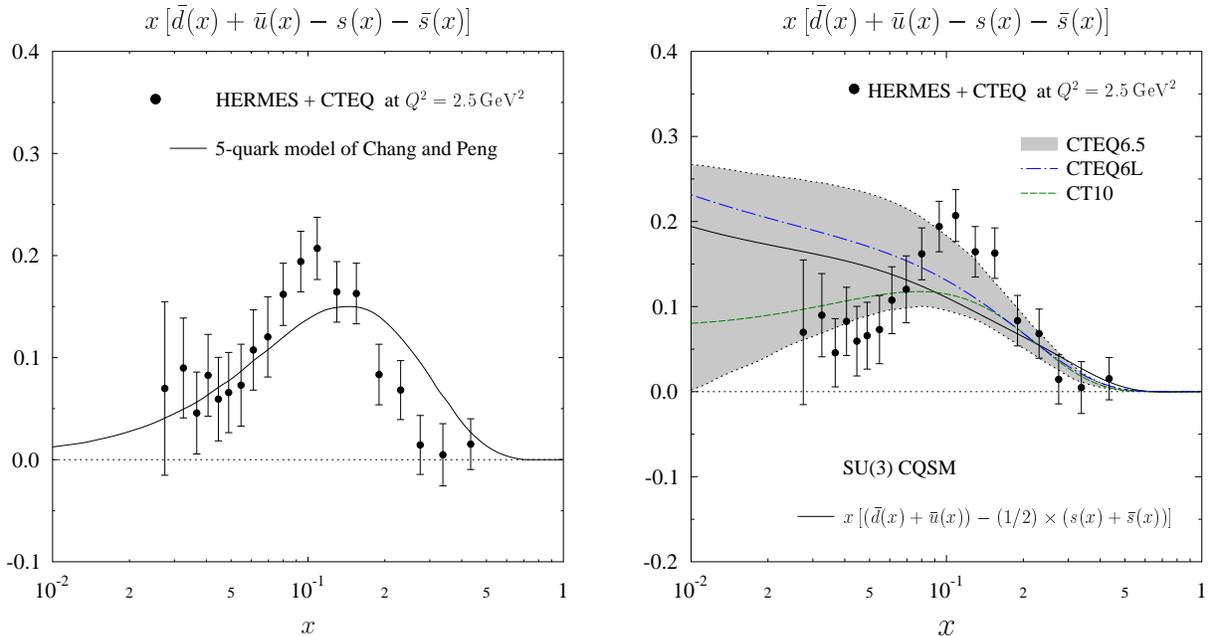}
\caption{The distribution $x \,[\bar{u}(x) + \bar{d}(x) - s(x) - \bar{s}(x)]$
obtained by using the CTEQ6.6 global fit for $\bar{u} + \bar{d}$ \cite{CTEQ08}
and the HEREMES SIDIS data for $s + \bar{s}$ \cite{HERMES08}.
The left panel shows the comparison with the prediction of the
BPHS 5-quark model due to Chang and Peng \cite{CP11B}, while the right panel
does the comparison with the prediction of the SU3 CQSM as well as
some other global fits \cite{CTEQ08},\cite{CT10}.}
\label{Fig09_xdbubssb_log}
\end{center} 
\end{figure}

Chang and Peng pushed their idea of ``intrinsic'' sea still
forward by considering the combination of the distributions
$\bar{u}(x) + \bar{d}(x) -  s(x) - \bar{s}(x)$ \cite{CP11B}.
According to them, this combination of the
distribution is particularly interesting, because the contribution
from the ``extrinsic'' seas is
expected to just cancel in this combination, so that it is only
sensitive to the ``intrinsic'' sea. Their
analysis then goes as follows. First, they proposed to extract
this distribution in an empirical way, i.e. by using
the HERMES SIDIS data for $x \,[s(x) + \bar{s}(x)]$ at 
$Q^2 = 2.5 \,\mbox{GeV}^2$ \cite{HERMES08} and the CTEQ6.6 fit for
the distribution $x\,[\bar{u}(x) + \bar{d}(x)]$ at the same
scale \cite{CTEQ08} as
\begin{equation}
 x \,[\bar{u}(x) + \bar{d}(x) - s(x) - \bar{s}(x)]  \  \Rightarrow \ 
 x \,[\bar{u}(x) + \bar{d}(x)]_{\rm CTEQ6.6} \ - \ 
 x \,[s(x) + \bar{s}(x)]_{\rm HERMES} .
\end{equation}
The resultant distribution is plotted by the filled circles in
the left panel of Fig.\ref{Fig09_xdbubssb_log}. A prominent
feature of the so-obtained
$x \,[\bar{u}(x) + \bar{d}(x) - s(x) - \bar{s}(x)]$ appears to have
an expected valence-like peaked structure.
Next, they calculated the corresponding distribution
on the basis of the BHPS model \cite{BHPS80},\cite{BPS81}, which gives
\begin{equation}
 \bar{u}(x) + \bar{d}(x) - s(x) - \bar{s}(x) \ = \ 
 P^{u \bar{u}} (x_{\bar{u}}) \ + \ P^{d \bar{d}} (x_{\bar{d}}) \ - \ 2 \,P^{s \bar{s}} (x_{\bar{s}}) .
\end{equation}
where $P^{Q \bar{Q}} (X_{\bar{Q}})$ is the $x$ distribution of $\bar{Q}$ in the
Fock component $| u u d Q \bar{Q} \rangle$ of the nucleon state vector.
In this calculation,
they assumed that the probability of the intrinsic sea is proportional to $1 / m^2_Q$ with
$m_Q$ being the mass of quark (antiquark) $Q$.
This BHPS prediction is then evolved to $Q^2 = 2.5 \,\mbox{GeV}^2$ by
taking $Q^2_{ini} = (0.5 \,\mbox{GeV})^2$ as the initial energy scale of evolution.
The answer is shown by the
solid curve in the left panel of Fig.\ref{Fig09_xdbubssb_log}. 
Chang and Peng emphasized that the qualitative agreement between
the data and the calculation provides strong supports to the existence of the
intrinsic $u$ and $d$ quark seas and also to the adequacy of the BHPS idea.

We point out that the valence-like peaked structure of the empirically
extracted distribution
$x \,[\bar{u}(x) + \bar{d}(x) - s(x) - \bar{s}(x)]$ may critically depend
on the following two factors, 
\begin{itemize}
\item the two-component structure of the HERMES SIDIS data for $x \,[s(x) + \bar{s}(x)]$ :
\item the relative magnitudes of the sea-like components of the $\bar{u} + \bar{d}$
distribution and the $s + \bar{s}$ distribution in the lower $x$ domain.
\end{itemize}
To confirm it, we first check what happens if we do not use the HERMES SIDIS data for
the strange quark distribution. In fact, from the viewpoint of internal consistency,
it would be more legitimate to extract the distribution in question
by using the same set of extraction framework for both of
$x\,[\bar{u}(x) + \bar{d}(x)]$ and $x \,[s(x) + \bar{s}(x)]$.
The dotted and dash-dotted curves in the right panel of Fig.\ref{Fig09_xdbubssb_log}
respectively correspond to the results
obtained by using the CTEQ6L fit \cite{CTEQ08} and the CT10 fit \cite{CT10}.
One finds a big difference between
these two fits. The distribution obtained from the CT10 fit has a peaked
structure, whereas
that obtained from the CTEQ6L does not. The origin of this difference can
primarily be
traced back to the relative magnitudes of the $s + \bar{s}$ and
$\bar{u} + \bar{d}$ distributions in the lower $x$ region.
(We point out that the CT10 fit gives larger magnitude of $s +\bar{s}$
distribution than the
CTEQ6L fit and also the NNPDF fit.) For reference, we also show in 
the right panel of Fig.\ref{Fig09_xdbubssb_log} the prediction of
the CQSM by the long-dashed curve. By the reason explained before, we have used the
reduced prediction for the $x [s(x) + \bar{s}(x)]$ distribution. It is interesting to see that the
resultant distribution $x \,[\bar{u}(x) + \bar{d}(x) - s(x) - \bar{s}(x)]$ is remarkably similar
in shape to that of the CTEQ6L fit, which does not show a peaked structure.
In any case, all these theoretical and semi-empirical predictions for the distribution
$x \,[\bar{u}(x) + \bar{d}(x) - s(x) - \bar{s}(x)]$ lie within the wide uncertainty band
indicated by the CTEQ6.5 fit, which in fact allows both of the peaked and
peak-less structures. Undoubtedly, to get more confident information
on the $x$-dependence
of this interesting combination $x\,[\bar{u}(x) + \bar{d}(x) - s(x) - \bar{s}(x)]$,
we need to get more reliable separate information for 
the light-flavor sea-quark distribution $x \,[\bar{u}(x) + \bar{d}(x)]$ and
the strange quark distribution $x\, [s(x) + \bar{s}(x)]$ in the nucleon.

An interesting idea of ``two component'' quark sea was also advocated
by Liu and the collaborators based on the path-integral formulation or
within the framework of the lattice QCD \cite{LCCP12},\cite{LD94},\cite{Liu00}.
According to them, the light-flavor $u$- and $d$-quark seas consist of the
connected sea and the disconnected sea, while the strange as well as charm
sea comes only from disconnected sea. On the basis of this idea, they carried
out a phenomenological extraction of the connected and disconnected
pieces of the light-flavor sea, $\bar{u}(x) + \bar{d}(x)$.
They first assume that the
disconnected sea component of the $\bar{u}(x) + \bar{d}(x)$ distribution is
proportional to the $s(x) + \bar{s}(x)$ distribution as
\begin{equation}
 \bar{u}^{ds} (x) \ + \ \bar{d}^{ds} (x) \ = \ 
 \frac{1}{R} \,\left[ s(x) \ + \ \bar{s}(x) \right] , 
\end{equation}
with the proportional constant
\begin{equation}
 R \ = \ \frac{\langle x \rangle_{s + \bar{s}}}
 {\langle x \rangle_{u + \bar{u} \,\,(DI)}} \ = \ 
 \frac{\langle x \rangle_{s + \bar{s}}}
 {\langle x \rangle_{\bar{u}^{ds} + \bar{u}^{ds}}} \ = \ 
 0.857(40) .
\end{equation}
which they estimated from lattice data. 
Then, they extracted the connected sea (CS)
component of the $\bar{u}(x) + \bar{d}(x)$ distribution, by using the CT10 PDF
fit for $\bar{u}(x) + \bar{d}(x)$ and the HERMES SIDIS data for the
strange quark distribution $s(x) + \bar{s}(x)$ as
\begin{eqnarray}
 \bar{u}^{cs} (x) \, + \, \bar{d}^{cs} (x) &\equiv&
 [\bar{u}(x) \, + \, \bar{d}(x)] \ - \ [\bar{u}^{ds}(x) \,+ \bar{d}^{ds}(x)]
 \nonumber \\
 &=& [\bar{u}(x) \, + \, \bar{d}(x)]_{\rm CT10} \ - \ 
 \frac{1}{R} \,[s(x) \,+ \,\bar{s}(x)]_{\rm HERMES} .
\end{eqnarray}

\vspace{5mm}
\begin{figure}[ht]
\begin{center}
\includegraphics[width=9.0cm]{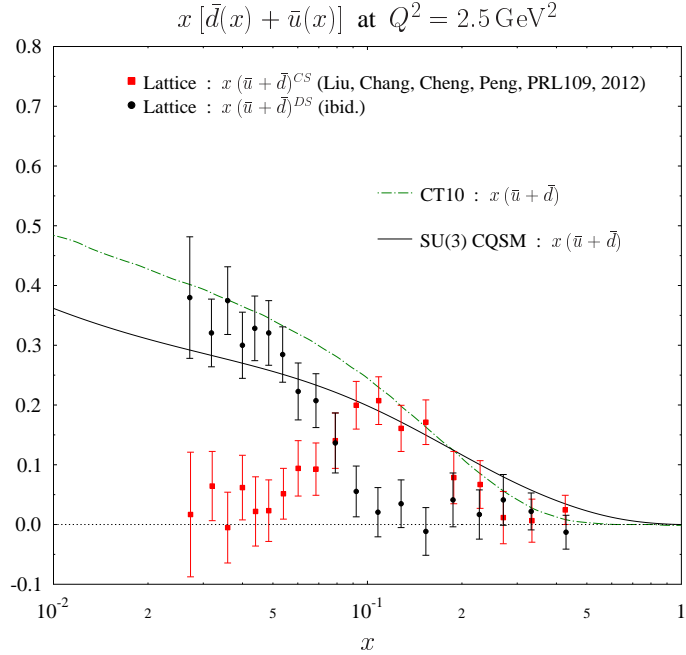}
\caption{The connected (filled circles) and disconnected (filled squares)
seas for $\bar{u}(x) + \bar{d}(x)$
extracted from CT10 PDF fit for $\bar{u}(x) + \bar{d}(x)$ \cite{CT10}
and the HERMES SIDIS data for the strange quark distribution
$s(x) + \bar{s}(x)$ \cite{HERMES08}, under the assumption that the
disconnected sea component of the $\bar{u}(x) + \bar{d}(x)$ distribution is
proportional to the $s(x) + \bar{s}(x)$ distribution \cite{LCCP12}.}
\label{Fig10_xdb_ub_log}
\end{center} 
\end{figure} 

The connected and disconnected seas for $\bar{u} + \bar{d}$ extracted
from the above-explained phenomenological analysis are shown
in Fig.\ref{Fig10_xdb_ub_log}
by the filled square (red in color) and the filled circles (black in color),
respectively.
Note that, by construction, the sum of these two components, i.e. the
connected sea and the disconnected sea, coincides with the CT10 global
fit shown by the dash-dotted curve.
They emphasize that the connected sea component so extracted appears to
have a valence-like peak around $x \simeq (0.1-0.2)$. However, it seems
to us that their separation into the two components
is not also independent on the structure of the HERMES data for the
strange quark distribution.    
Going back to the original physical idea of Liu et al. \cite{LD94},\cite{Liu00},
it is certainly true that the distribution $\bar{u}(x) + \bar{d}(x)$
is generally given as a sum of the connected and disconnected sea
contributions. This reminds us of the similar idea of Chang and
Peng \cite{CP11A},\cite{CP11B}.
In their language, it roughly corresponds to saying that the distribution
$\bar{u}(x) + \bar{d}(x)$ consists of ``intrinsic'' sea and ``extrinsic''
sea.
However, it should be recognized that there is no
rigorous correspondence between the two terminologies, i.e. the idea
of ``intrinsic'' and ``extrinsic'' seas and that of connected and
disconnected seas in the language of the lattice QCD.
In fact, according to Chang and Peng, the strange quark distribution
$x \,[s(x) + \bar{s}(x)]$ also contains the ``intrinsic'' component.
But, this ``intrinsic'' sea requires at least 5-quark component,
which needs to take account of disconnected seas within the
framework of lattice QCD.

In our opinion, just like that the decomposition into the ``intricsic''
sea and the ``intrinsic'' is a model-dependent idea, the decomposition into
the connected sea and the disconnected sea is an idea, which has
a definite meaning only within the framework of the lattice QCD
formulation of QCD.
Model independent notion is only the separation into a quark and
anti-quark distributions. In fact, we show in Fig.\ref{Fig10_xdb_ub_log}
also the prediction of the SU(3) CQSM for $x \,[\bar{u}(x) + \bar{d}(x)]$
by the solid curve.
Although it slightly underestimates the magnitude of $\bar{u} + \bar{d}$
in the small $x$ region as compared with the CT10 global fit \cite{CT10},
an important fact is that it does not show any two-component
structure just like the CT10 fit does not.
Furthermore, within the framework of the CQSM, there is no idea
of decomposing the anti-quark distributions into the two components like
the ``intrinsic'' and ``extrisic'' sea. Both are contained within a
single theoretical scheme without any separation between them.
This reconfirms again that the separation of the anti-quark distribution
into the ``intricsic'' and ``extrinsic'' components or into the
connected and disconnected seas is a {\it theoretical-scheme-dependent idea},  
although we never deny its usefulness for understanding the nature
quark seas in the nucleon.


\section{Flavor SU(3) CQSM and longitudinally polarized PDFs}

In this section, we compare the predictions of the SU(3) CQSM for the
longitudinally-polarized PDFs with empirically extracted ones.
Similarly as for the unpolarized PDFs, to get a feeling about the
degrees of success or failure of the model, we first compare our
predictions with the recently reported global fits of the longitudinally
polarized PDFs by the NNPDF group, i.e. NNPDFpol1.0 \cite{NNPDFpol1.0}.
The NNPDF fits for the longitudinally polarized PDFs are given at
$Q^2 = 1 \,\mbox{GeV}^2$ for the following combinations of the PDFs : 

\begin{itemize}

\item the flavor singlet, $\Delta \Sigma (x) \equiv
\sum_{i=1}^{n_f} \,(\Delta q_i (x) + \Delta \bar{q}_i (x))$,

\item the gluon, $\Delta g(x)$,

\item the isospin triplet, $\Delta T_3 (x) \equiv
(\Delta u(x) + \Delta \bar{u} (x)) - (\Delta d(x) + \Delta \bar{d}(x))$,

\item the SU(3) octet, $\Delta T_8 (x) \equiv
\Delta u(x) + \Delta \bar{u}(x) + \Delta d(x) + \Delta \bar{d}(x)
- 2 \,(\Delta s(x) + \Delta \bar{s}(x))$,

\item the $u$ plus $\bar{u}$, $\Delta u(x) + \Delta \bar{u}(x)$,

\item the $d$ plus $\bar{d}$, $\Delta d(x) + \Delta \bar{d}(x)$,

\item the $s$ plus $\bar{s}$, $\Delta s(x) + \Delta \bar{s}(x)$.

\end{itemize}

For making a comparison, the CQSM predictions are evolved from
the initial scale $Q^2_{ini} = 0.30 \,\mbox{GeV}^2$ to the
scale $Q^2 = 1.0 \,\mbox{GeV}^2$ where the NNPDF fits are given.
As before, the gluon distribution at the initial scale is
simply set to be zero.

\begin{figure}[ht]
\begin{center}
\includegraphics[width=15.0cm]{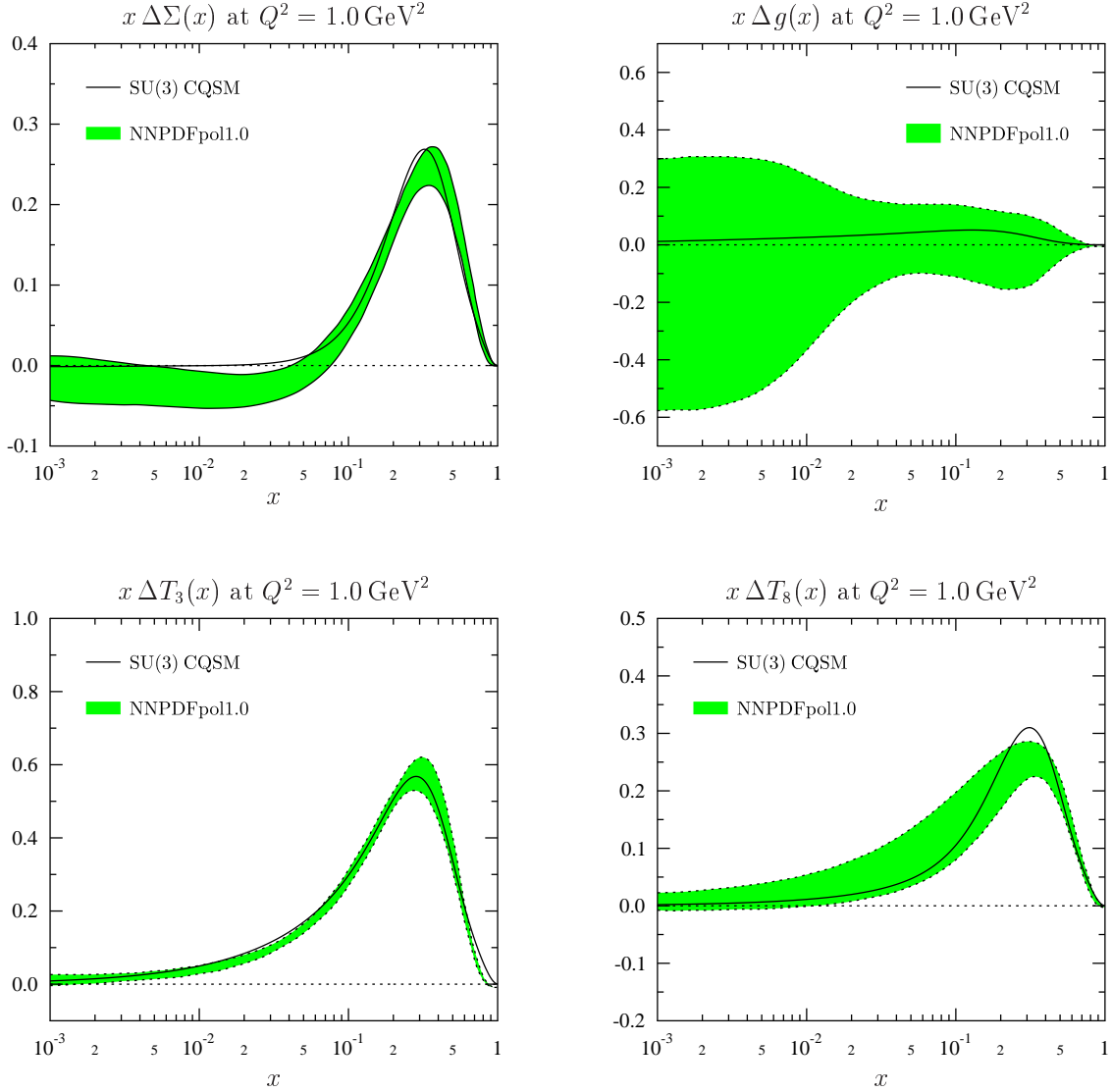}
\caption{The NNPDFpol1.0 fits for the polarized PDFs
$x \,\Delta T_3 (x)$, $x \,\Delta T_8 (x)$, $x \,\Delta \Sigma (x)$,
and $x \,\Delta g(x)$ in comparison with the predictions of
the SU(3) CQSM.}
\label{Fig11_AllxDel1}
\end{center} 
\end{figure} 

The Fig.\ref{Fig11_AllxDel1} shows the comparison for the polarized PDFs,
$x \,\Delta T_3 (x)$, $x \,\Delta T_8 (x)$, $x \,\Delta \Sigma (x)$,
and $x \,\Delta g(x)$.
One can say that the agreement between the theoretical predictions and
the global fits is fairly good, obviously much better than the case of the
unpolarized PDFs. From the similar analysis for the
unpolarized distributions, we could have expected a good agreement for
the non-singlet distributions like $\Delta T_3 (x)$ and $\Delta T_8 (x)$.
However, different from the unpolarized case, we clearly get
much better agreement also for the
flavor-singlet quark distribution $\Delta \Sigma (x)$ and the
gluon distribution $\Delta g (x)$.
Remember that, in the case of unpolarized PDFs, the flavor
singlet quark distribution was not reproduced very well.
We argued that a possible reason of this discrepancy
may be traced back to the neglect of the fact that the quark fields
is likely to carriy only about 80 \% of the nucleon momentum at the
model scale of $Q^2_{ini} \simeq 0.30 \,\mbox{GeV}^2$.
Put another way, a good agreement for the flavor-singlet polarized
distribution $\Delta \Sigma (x)$ indicates that the neglect of the gluon
contribution at the model energy scale does little harm.
Accordingly, the following picture emerges. At the low energy
scale corresponding to the CQSM, the gluon is likely to carry about 20 \%
of the nucleon momentum fraction, but it carries negligiblly small
polarization.
It is interesting to point out that this observation is consistent with
the claim in the paper by Efremov, Goeke, and Pobylitsa \cite{EGP00}.
In fact, on the general ground of large $N_c$ QCD, they argued that the
polarized gluon distribution is $1 / N_c$ suppressed compared to the
unpolarized one. 

Note however that this does not means
that the gluon polarization remains to be small also at the high
energy scales. If the quark has a positive polarization at the
low energy scale, the polarization of gluon grows rapidly
through the process of scale evolution. To get a feeling of
the evolution effect, we solve the coupled evolution equation at the NLO
for the net quark polarization and the gluon polarization at the NLO
by starting with the initial condition of the CQSM,
i.e. $\Delta \Sigma = 0.35$ and $\Delta G = 0.0$ at
$Q^2_{ini} = 0.30 \,\mbox{GeV}^2$. The net gluon polarization
$\Delta G$ obtained in this way is shown in Table \ref{Table:Gluon_spin}
for some typical values of $Q^2$.

\vspace{3mm}
\newcommand{\lw}[1]{\smash{\lower2.ex\hbox{#1}}}
\begin{table}[h]
\caption{The net longitudinal gluon polarization $\Delta G$ in dependence of
$Q^2$, obtained by solving the QCD evolution equation at the NLO
under the assumption that $\Delta G = 0$ and $\Delta \Sigma = 0.35$ at the
initial energy scale of the CQSM.}
\label{Table:Gluon_spin}
\vspace{2mm}
\begin{center}
\renewcommand{\arraystretch}{1.0}
\begin{tabular}{ccccc}
\hline\hline
 $Q^2 \,[\mbox{GeV}^2]$ \ \ \ \ & \ \ \ \ 0.30 \ \ \ \ & \ \ \ \ 1.0 
 \ \ \ \ & \ \ \ \ 4.0 \ \ \ \ & \ \ \ \ 10.0 \ \ \ \ \\
 \hline
 $\Delta G (Q^2)$ \ \ \ \ & \ \ \ \ 0.0 \ \ \ \ & \ \ \ \ 0.21 
 \ \ \ \ & \ \ \ \ 0.40 \ \ \ \ & \ \ \ \ 0.51 \ \ \ \\
\hline \hline
\end{tabular}
\end{center}
\end{table}

We see that, even if we assume that $\Delta G = 0$ at the low energy
model scale, the gluon polarization increases rapidly as $Q^2$ becomes
large. Very recently, the DSSV collaboration carried out a systematic analysis
of the gluon polarization in the nucleon by paying particular attention
to the data offered by polarized proton-proton collision available
at the BNL Relativistic Heavy Ion Collider (RHIC) \cite{DSSV14}. 
The final answer of their new global fit, corresponding to the scale
$Q^2 = 10 \,\mbox{GeV}^2$, is shown in Fig.5 of their paper.
This figure give estimates for the 90 \% C.L. area in the plane spanned
by the truncated moments of $\Delta g(x)$ calculated in 
$0.05 \le x \le 1$ and $0.001 \le x \le 0.05$.
Their result can be summarized as
\begin{equation}
 \int_{0.05}^1 \,\Delta g(x) \,dx \ = \ 0.194
 \begin{array}{l}
 \raisebox{-0.4ex}[0pt]{+ \,0.060} \\
 \raisebox{+0.4ex}[0pt]{\,-- \,\,0.060} \\
 \end{array} \ .
\end{equation}
and
\begin{equation}
 \int_{0.001}^{0.05} \,\Delta g(x) \,dx \ = \ 0.166
 \begin{array}{l}
 \raisebox{-0.4ex}[0pt]{+ \,0.062} \\
 \raisebox{+0.4ex}[0pt]{\,-- \,\,0.046} \\
 \end{array} \ .
\end{equation}
Summing up the contributions of both $x$ region, this would give
\begin{equation}
 \int_{0.001}^{1} \,\Delta g(x) \,dx \ = \ 0.361
 \begin{array}{l}
 \raisebox{-0.4ex}[0pt]{+ \,0.683} \\
 \raisebox{+0.4ex}[0pt]{\,-- \,\,0.522} \\
 \end{array} \ .
\end{equation}
Note that the central value of the moment of $\Delta g(x)$, i.e., $\Delta G$
is positive with sizable magnitude, although the negative value is not
completely excluded due to still large uncertainty coming from the integral
in the small $x$ region.
It is clear from the analysis above that positive value of
$\Delta G$ is theoretically more than natural.
If the result of global fits at high energy scale give negative gluon
polarization, it would rather give us a puzzle, which is difficult to solve.

\vspace{4mm}
\begin{figure}[ht]
\begin{center}
\includegraphics[width=15.0cm]{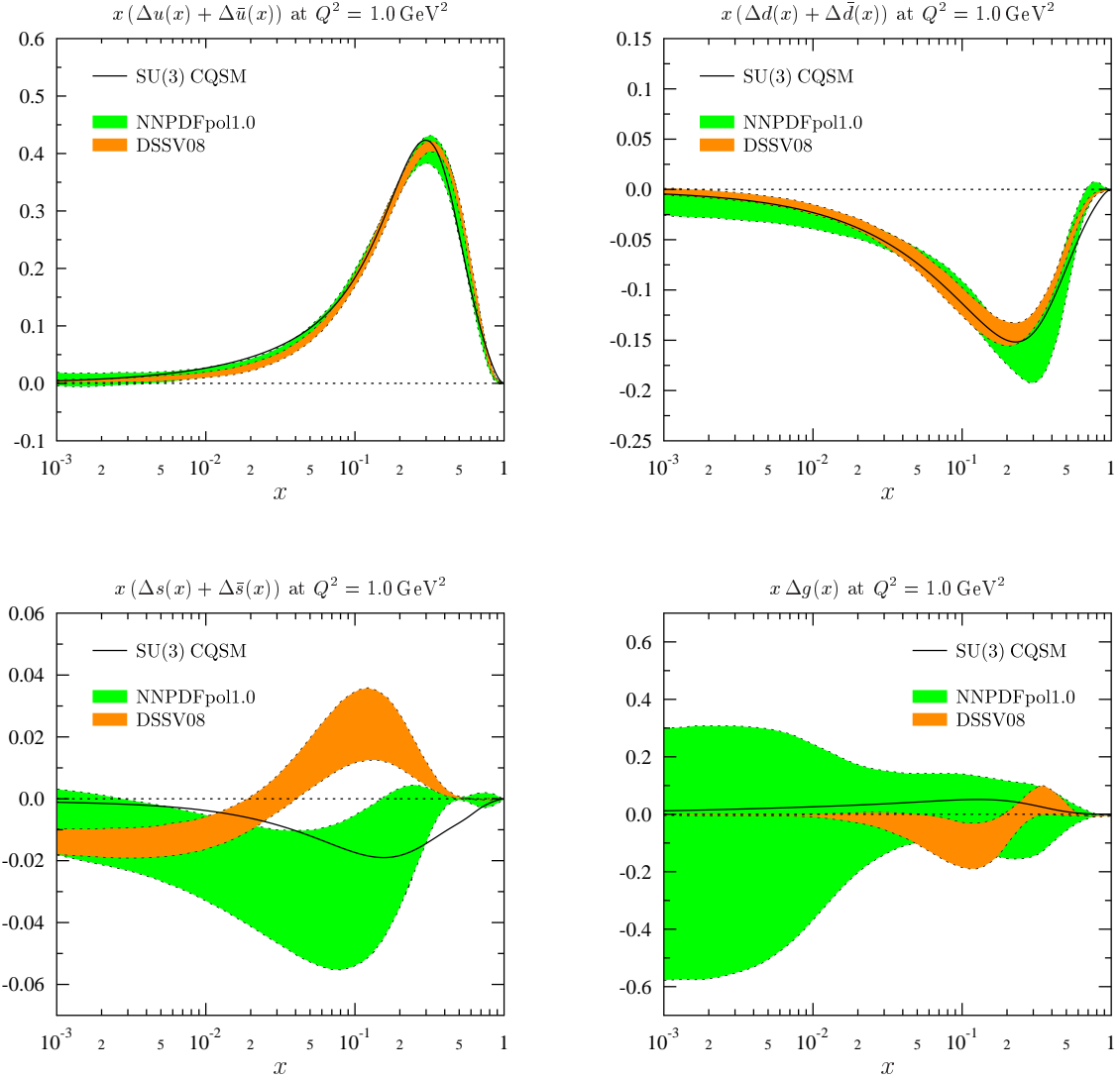}
\caption{The SU(3) CQSM predictions for the distributions
$x \,(\Delta u(x) + \Delta \bar{u}(x))$,
$x \,(\Delta d(x) + \Delta \bar{d}(x))$,
$x \,(\Delta s(x) + \Delta \bar{s}(x))$, and also for
$x \,\Delta g(x)$ in comparison with the NNPDFpol1.0 fits \cite{NNPDFpol1.0}
together with the DSSV08 fits \cite{DSSV08}.}
\label{Fig12_AllxDel2}
\end{center} 
\end{figure} 

Next, in Fig.\ref{Fig12_AllxDel2}, we compare the prediction of the SU(3) CQSM for
the distributions $x \,(\Delta u(x) + \Delta \bar{u}(x))$,
$x \,(\Delta d(x) + \Delta \bar{d}(x))$,
$x \,(\Delta s(x) + \Delta \bar{s}(x))$, and also for
$x \,\Delta g(x)$ with the NNPDF fits together with a little older
DSSV08 global fits.
One sees that the model predictions for $x \,(\Delta u(x) + \Delta \bar{u}(x))$
and $x \,(\Delta d(x) + \Delta \bar{d}(x))$ are remarkably consistent with
both of the NNPDF fits and the DSSV08 fits \cite{DSSV08},
which are close to each other,
anyway. What is problematical is the polarized strange quark distribution.
The NNPDF fit gives negative strange quark polarization, whereas the
DSSV08 fit gives positive polarization at least in the larger $x$ region.
This means that the presently-available empirical information is not
sufficient enough to
determine the longitudinally polarized strange quark distribution with confidence.
The difference between the two determinations lies in the fact
that the DSSV fits depends more heavily on the data of semi-inclusive
DIS reactions. As we have already pointed out, we feel that our understanding
of the mechanism of the semi-inclusive reactions has not reached a satisfactory
level as compared with the inclusive DIS reactions. At any rate,
it is interesting to point out that the prediction of the SU(3) CQSM
for the strange quark polarization is negative and consistent with
the NNPDF fit at least qualitatively. 
Finally, the shapes of the gluon distributions are also fairly
different between the NNPDF fit and the DSSV08 fit. However, the
uncertainty bands for $\Delta g(x)$ is sizably large in both fits.
We point out that the prediction of the CQSM, obtained by assuming
$\Delta g(x) = 0$ at the model scale, is within the (broad) error band
of the NNPDF fit. 

\vspace{3mm}
\begin{figure}[ht]
\begin{center}
\includegraphics[width=10.0cm]{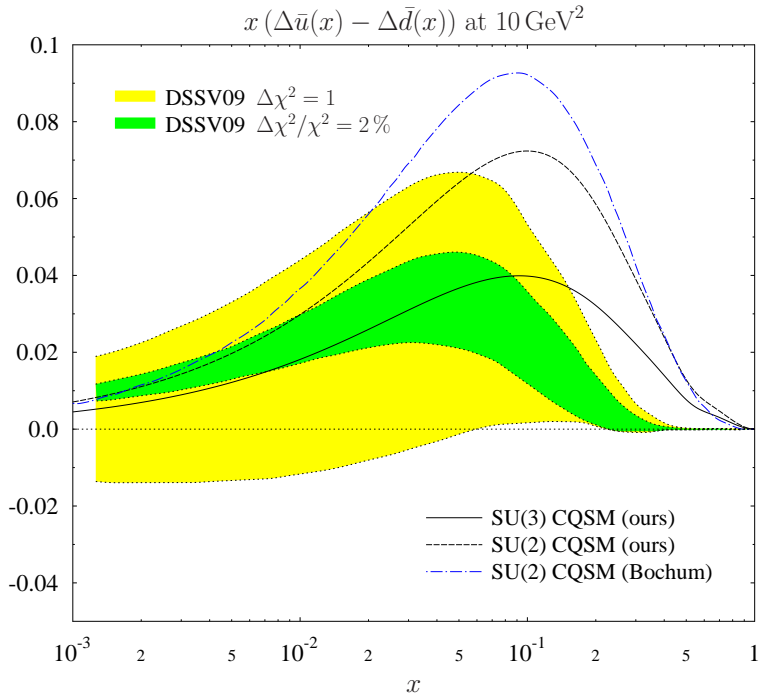}
\caption{The predictions of the SU(2) and SU(3) CQSM for the
polarized light-flavor sea-quark asymmetry
$x \,(\Delta \bar{u}(x) - \Delta \bar{d})$ in comparison with
the DSSV09 global fit \cite{DSSV09}.
The dash-dotted curve is the prediction
of the SU(2) CQSM by the Bochum group \cite{DGPW00},
whereas the long-dashed
curve is our prediction in the same model. The prediction of
the SU(3) CQSM is shown by the solid curve.}
\label{Fig13_xDelubmdb}
\end{center} 
\end{figure} 

After confirming that the predictions of the SU(3) CQSM for the longitudinally
polarized PDFs are remarkably consistent with the empirically extracted
PDFs, especially the NNPDFpol1.0 fit, we now turn to more detailed
inspection of the flavor structure of the longitudinally polarized
sea-quark (anti-quark) distributions. 
We first show in Fig.\ref{Fig13_xDelubmdb}
the predictions of the CQSM for the flavor (or isospin)
asymmetry for the longitudinally polarized sea-quark distributions, i.e.,
$x \,(\Delta \bar{u}(x) - \Delta \bar{d}(x))$ in comparison with the
DSSV09 fit. The thinner (yellow in color) shaded area and the thicker
(green in color) shaded
area are the allowed bands of the DSSV09 fit given at
$Q^2 = 10 \,\mbox{GeV}^2$, respectively with
$\Delta \chi^2 \,/\,\chi^2 = 2 \%$ and with $\Delta \chi^2 = 1$.  
The solid curve is the prediction of the SU(3) CQSM, while
the dashed curve is that of the SU(2) CQSM. The corresponding prediction
of the Bochum group within the SU(2) CQSM is also shown for
reference \cite{DGPW00}.
We first point out that our prediction and the that of Bochum group are
sizably different in spite that they are the predictions based on the same
SU(2) CQSM. The reason of this discrepancy is not absolutely clear. 
We conjecture that a possible reason is that their calculation use schematic
soliton profile function, while we use the solution of the self-consistent mean-field
equation.
Another reason may be that their predictions were obtained by using
what-they-call the ``interpolation formula'', which is an approximate
method of calculating PDFs or any nucleon observables within the
framework of the CQSM \cite{DPPPW96}.
In any case, our prediction for $\Delta \bar{u}(x) - \Delta \bar{d}(x)$
is significantly smaller than that of the Bochum group.   
Furthermore, the prediction of the SU(3) CQSM is much smaller than that
of the SU(2) model. This provides a rare case in which the SU(3) CQSM
and the SU(2) CQSM give a significantly different prediction for the
light-flavor $u$- and $d$-quark distributions.
One can see that the prediction of the SU(3) CQSM is order of magnitude
consistent with the DSSV fit, although the positions of peak are
slightly different. At any event, we find that the CQSM predicts fairly large flavor
(isospin) asymmetry not only for the unpolarized sea-quark distributions
but also for the longitudinally polarized sea-quark distributions.
This should be contrasted with the prediction of the meson cloud models.
Although it is known that the meson cloud models nicely reproduce the
isospin asymmetry of the unpolarized sea-quark distributions, their
predictions for the longitudinally polarized sea-quark distributions
are generally very small or even diverging.  

The reason why $\Delta \bar{u}(x) - \Delta \bar{d}(x)$ is large was
already discussed in several papers by Diakonov
et al. \cite{DPPPW96},\cite{DPPPW97}.
According to their large-$N_c$ argument, $u(x) - d(x)$ and also
$\bar{u}(x) - \bar{d}(x)$ are $1 / N_c$ suppressed as compared with
$\Delta u(x) - \Delta d(x)$ and $\Delta \bar{u}(x) - \Delta \bar{d}(x)$,
which was claimed to explain the fact that $\Delta \bar{u}(x) - \Delta \bar{d}(x)$
is large. However, in reality $N_c = 3$ and the explicit numerical calculation
within the CQSM reveals that $\bar{u}(x) - \bar{d}(x)$ and
$\Delta \bar{u}(x) - \Delta \bar{d}(x)$ actually have comparable
magnitudes \cite{Wakamatsu03A}.
Furthermore, the large-$N_c$ argument tells us little
about the $x$-dependencies of these distribution functions.
The explicit $x$-dependencies can be known only through explicit
numerical calculation within the CQSM.
To answer the above question beyond the simple large-$N_c$
counting argument, we therefore think it instructive to look more
closely at the predictions of the CQSM for four basic twist-2 PDFs,
i.e. the isoscalar and isovector combinations of the unpolarized and
longitudinally polarized PDFs.
(To avoid inessential complexity, we show here the predictions of
the SU(2) CQSM. This is enough because the essential physics of
strong spin-isospin correlation is already embedded in the SU(2)
model in the form of rotational symmetry-breaking mean-field.)

\vspace{5mm}
\begin{figure}[ht]
\begin{center}
\includegraphics[width=14.0cm]{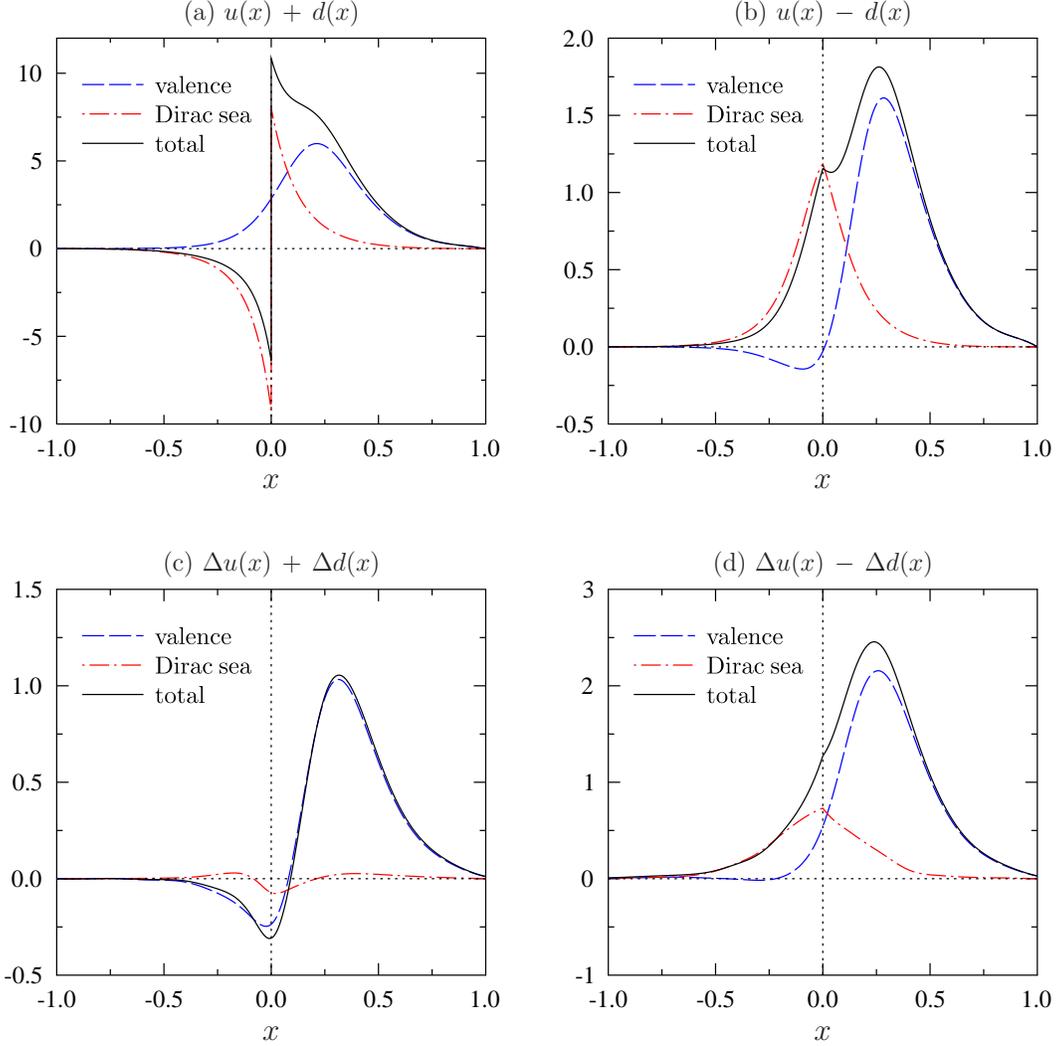}
\caption{The predictions of the SU(2) CQSM for the four basic twist-2 PDFs,
i.e. the isoscalar and isovector combinations of the unpolarized and
longitudinally polarized PDFs. In these figures, the long-dashed curves
(blue in color) stand for the contributions of the three valence quarks
in the mean field, whereas the dash-dotted curves (red in color) are those
of the vacuum-polarized Dirac-sea quarks. The sums of these two contributions
are shown by the solid curves.}
\label{Fig14_twist2PDF_SU2}
\end{center} 
\end{figure}

In Fig.\ref{Fig14_twist2PDF_SU2}, the long-dashed curves (blue in color)
stand for the contributions of the three valence quarks in the mean field,
whereas the dash-dotted curves (red in color) are those of the
vacuum-polarized Dirac-sea quarks. The sums of these two contributions
are shown by the solid curves.
In these figures, the distribution functions in the negative $x$ region
must be interpreted as antiquark distributions according to the rule : 
\begin{eqnarray}
 u(-x) \ + \ d(-x) &=& - \,[\bar{u}(x) \ + \ \bar{d}(x)], \label{Eq:CC1} \\
 u(-x) \ - \ d(-x) &=& - \,[\bar{u}(x) \ - \ \bar{d}(x)], \label{Eq:CC2} \\ 
 \Delta u(-x) \ + \ \Delta d(-x) &=& 
 + \,[\Delta \bar{u}(x) \ + \ \Delta \bar{d}(x)], \label{Eq:CC3} \\
 \Delta u(-x) \ - \ \Delta d(-x) &=& 
 + \,[\Delta \bar{u}(x) \ - \ \Delta \bar{d}(x)], \label{Eq:CC4}
\end{eqnarray}
with $0 < x < 1$. The sign-difference between the unpolarized and
longitudinally polarized distributions originates from the symmetry
properties of those under the charge-conjugation transformation.
As one can see, the contributions of the three valence quarks have
more or less similar shapes. They are peaked around $x \sim (0.2 - 0.4)$.
On the other hand, one sees totally different behaviors of the
contributions of Dirac-sea quarks in different distribution functions,
all of which are already known to play important
roles in reproducing the empirical distributions.
One may however notice that the Dirac-sea contributions
are surprisingly similar in shape for the two isovector distributions, i.e.
for $u(x) - d(x)$ and $\Delta u(x) - \Delta d(x)$.
The fact that $u(x) - d(x) > 0$ in the negative $x$
region means $\bar{u}(x) - \bar{d}(x) < 0$ for the physical value of $x$
in the range $0 < x < 1$, which naturally explains the famous NMC
observation. On the other hand, $\Delta u(x) - \Delta d(x) > 0$ in the
negative $x$ region dictates that $\Delta \bar{u}(x) - \Delta \bar{d}(x) > 0$
for physical $x$. 
We recall that, in the energy spectrum of the single-particle Dirac
equation for quarks under the hedgehog mean field, there are
two (deformed) Dirac continuums : the positive energy one
and the egative energy one. Here, let us concentrate on the negative
energy Dirac continuum and also on the Dirac-sea contribution to
the PDFs in the negative $x$ region, which correspond to
antiquark distributions. The strong similarity in the shapes of
$u(x) - d(x)$ and $\Delta u(x) - \Delta d(x)$ in the negative $x$ region
actually corresponds to anticorrelation, because of the rules
(\ref{Eq:CC2}) and (\ref{Eq:CC4}).
It appears that this anticorrelation is compatible with the grand spin zero
nature of negative energy Dirac continuum, although more
convincing argument is highly desirable.
(We recall here the fact that the mean-field solution under
the hedgehog potential is known to have a quantum number of $K = 0$,
where $\bm{K} \equiv \bm{S} + \bm{T}$, with $\bm{S}$ and $\bm{T}$ being
the ordinary spin and isospin operators, is called the grand spin operator.)   


\begin{figure}[ht]
\begin{center}
\includegraphics[width=15.0cm]{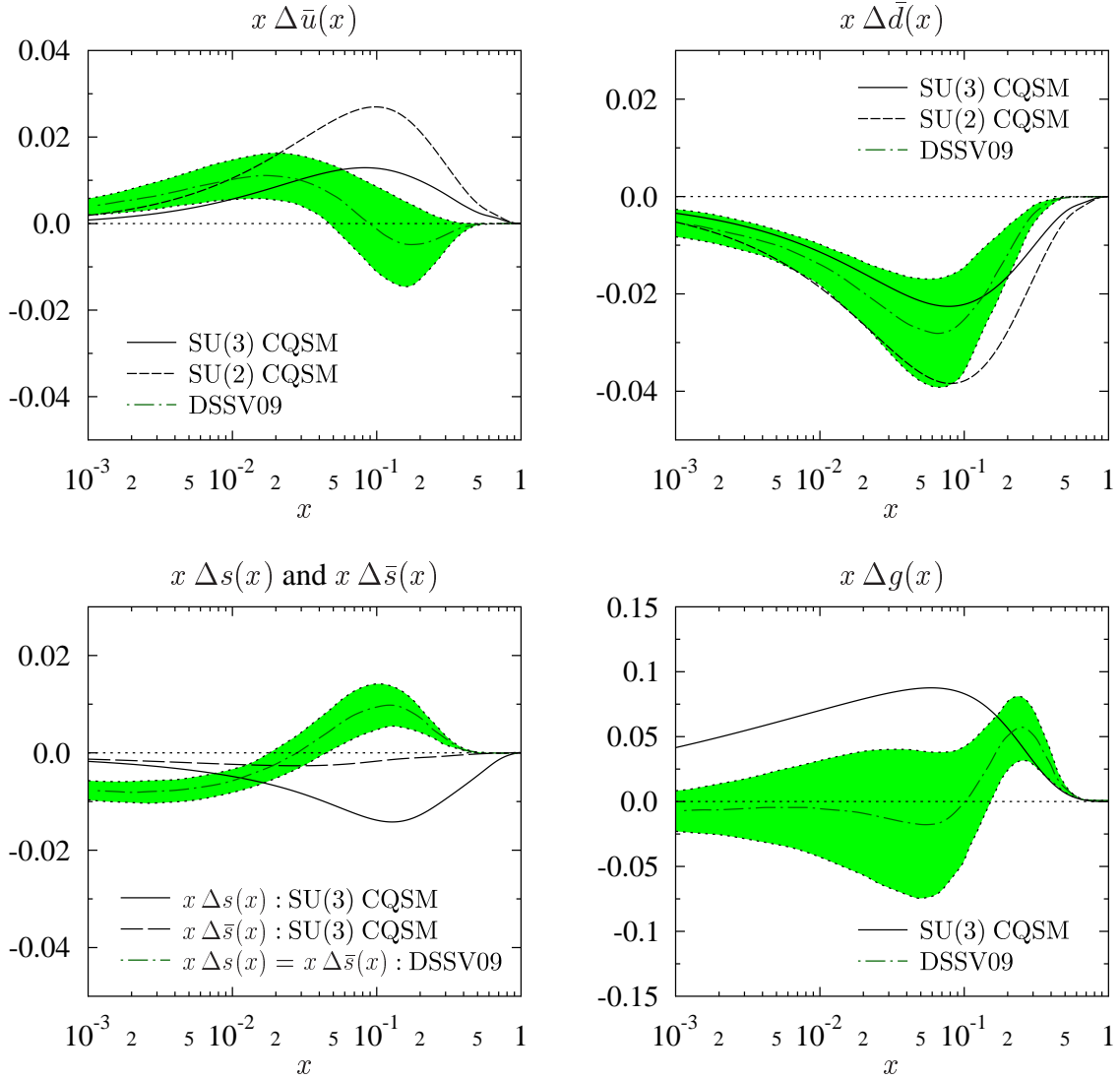}
\caption{The predictions of the SU(3) CQSM for the longitudinally
polarized sea-quark distributions $x \,\bar{u}(x)$, $x \,\bar{d}(x)$,
$x \,\Delta s(x)$, $x \,\bar{s}(x)$, and the polarized gluon
distribution $\Delta g(x)$ evolved to $Q^2 = 10 \,\mbox{GeV}^2$
in comparison with the DSSV09 global fits \cite{DSSV09}.}
\label{Fig15_xdelq_SU32}
\end{center} 
\end{figure} 

We have seen that the CQSM predicts large flavor asymmetry not only for
the unpolarized sea-quark distribution but also for the longitudinally
polarized sea-quark distribution, i.e.
$\Delta \bar{u}(x) - \Delta \bar{d}(x) > 0$.
Also interesting to know is separate information for polarized $\bar{u}$ and
$\bar{d}$ seas. Shown in Fig.\ref{Fig15_xdelq_SU32} are
the predictions of the SU(3) CQSM in comparison with the DSSV09 fits \cite{DSSV09}.
For completeness, we also show a comparison for the polarized strange
quark distribution and the polarized gluon distribution, because they
are also given as a set in the DSSV09 fits.
The model predicts that $x \,\Delta \bar{u}(x)$ is positive, while
$x \,\Delta \bar{d}(x)$ is negative with sizable magnitude.
One confirms that the predictions of the SU(3) CQSM for
$x \,\Delta \bar{u}(x)$ and for $x \,\Delta \bar{d}(x)$ are both
order of magnitude consistent with the DSSV09 fits. The DSSV central
fit for the polarized $\bar{u}$ distribution shows a nodal behavior
around $x \sim 0.08$, which is not reproduced by the CQSM.
From the theoretical viewpoint, however, such nodal behavior of the
distribution $\Delta \bar{u}(x)$ is difficult to understand.
We again suspect that our incomplete understanding of the semi-inclusive
processes can be a cause of this unnatural nodal behavior of the global fit.
Turning to the strange quark distributions, the SU(3) CQSM predicts
negative polarization, while the result of the DSSV09 fit is positive
in the higher $x$ range, where the distribution is dominant.
However, we have already pointed out that the more recent NNPDF fits
give negative strange quark polarization \cite{NNPDFpol1.0} in
qualitatively consistent with the prediction of the SU(3) CQSM.
Incidentally, in the DSSV
analysis, the equality of the polarized strange and anti-strange
distributions are assumed from the beginning.
Very interestingly, according to the SU(3) CQSM, the negative polarization
of the strange plus anti-strange distribution turns out to mostly come from
the strange quark and the polarization of the anti-strange quark is very small.
This means that the model predicts sizable particle-antiparticle asymmetry
not only for the unpolarized strange quark distributions but also for the
longitudinally polarized ones.     
We recall that this feature is also qualitatively consistent with
the picture of the kaon cloud model proposed by Signal and Thomas \cite{ST87},
Burkardt and Warr \cite{BW92}, and also by Brodsky and Ma \cite{BM96}.
In fact, the strange sea in the proton
is thought to be generated through the virtual dissociation
process of the proton into the $\Lambda$ and the $K^+$, i.e.
$p \rightarrow \Lambda + K^+$.
Note an apparent asymmetry of the $s$-quark and $\bar{s}$-quark in
this process. The $s$-quark is contained in the spin one-half $\Lambda$,
while the $\bar{s}$-quark is contained in the spin zero $K^+$.
This naturally explains the reason why the polarization of $\bar{s}$-quark
is smaller than that of $s$-quark.

\section{Flavor SU(3) CQSM and charge-symmetry-violating PDFs}

The effective lagrangian, which takes account of the 
charge-symmetry-violation (CSV), is given as
\begin{equation}
 {\cal L} \ = \ {\cal L}_0 \ + \ {\cal L}_{SB} \ + \ {\cal L}_{CSV}, 
\end{equation}
where ${\cal L}_0$ and ${\cal L}_{SB}$ were already given, while the CSV part
${\cal L}_{CSV}$ can be written as
\begin{equation}
 {\cal L} \ = \ - \,\Delta m \,\,\bar{\psi} \,\frac{\lambda_3}{2} \,\psi ,
\end{equation}
with $\Delta m \equiv m_u - m_d \simeq - \,4 \,\mbox{MeV}$.
The CSV effects for the PFDs in the nucleon can be investigated by
treating this SU(2) breaking part of the effective lagrangian as 
the first order perturbation. The general method is exactly the same
as the one used in the perturbative treatment
of the SU(3) symmetry breaking term ${\cal L}_{SB}$.
Note however mass difference between the $u$ and $d$ quarks
is far smaller than that between the strange quark and the $u,d$ quarks.
Consequently, the perturbative treatment of the CSV part ${\cal L}_{CSV}$
has even better foundation than that of ${\cal L}_{SB}$.
Since the necessary formalism is already explained in our previous paper,
we do not repeat the detailed derivation here. For completeness, however,
we just summarize below the final theoretical expressions, which
are necessary for the actual calculation.

Within the framework of the SU(3) CQSM, the PDFs for the $u, d, s$ quarks
are represented as linear combinations of three independent functions
$q^{(0)}(x)$, $q^{(3)}(x)$, and $q^{(8)}(x)$ as
\begin{eqnarray}
 u (x) \ &=& \ \frac{1}{3} \,q^{(0)} (x) \ + \ \frac{1}{2} \,q^{(3)} (x)
 \ + \ \frac{1}{2 \sqrt{3}} \,q^{(8)} (x) , \\
 d (x) \ &=& \ \frac{1}{3} \,q^{(0)} (x) \ - \ \frac{1}{2} \,q^{(3)} (x) 
 \ + \ \frac{1}{2 \sqrt{3}} \,q^{(8)} (x) , \\
 s (x) \ &=& \ \frac{1}{3} \,q^{(0)} (x) \hspace{24mm}
 \ - \ \frac{1}{\sqrt{3}} \,\,q^{(8)} (x). \label{s}
\end{eqnarray}
In the SU(3) symmetric limit, these three distributions generally consist of 
the zeroth order term and the first order term in the corrective
angular velocity $\Omega$ of the rotating soliton. (The zeroth order term
corresponds to the mean-field predictions.) They are given in the form : 
\begin{eqnarray}
 q^{(0)}(x) \ &=& \ \left\langle 1 \right\rangle_N \cdot f(x), \\
 q^{(3)}(x) \ &=& \ \left\langle \frac{D_{38}}{\sqrt{3}} \right\rangle_N
 \cdot f(x) \nonumber \\ 
 \ &+& \ \left\langle \sum_{i=1}^3 \,\left\{ D_{3i}, R_i \right\} \right\rangle_N
 \cdot k_1 (x) \ + \ 
 \left\langle \sum_{K=4}^7 \,\left\{ D_{3K}, R_K \right\} \right\rangle_N
 \cdot k_2 (x) \\
 q^{(8)}(x) \ &=& \ \left\langle \frac{D_{88}}{\sqrt{3}} \right\rangle
 \cdot f(x) \nonumber \\
 \ &+& \ \left\langle \sum_{i=1}^3 \,\left\{ D_{8i}, R_i \right\} \right\rangle_N
 \cdot  k_1 (x) \ + \ 
 \left\langle \sum_{K=4}^7 \,\left\{ D_{8K}, R_K \right\} \right\rangle_N
 \cdot k_2 (x) .
\end{eqnarray}
The functions $f(x)$, $k_1 (x)$, and $k_2 (x)$ are defined in Eqs.(33),
(76), and (77) of the paper \cite{Wakamatsu03B}.
(They are all calculable, once the
solutions of the mean-field equations are given.)
Here, the terms containing the function $f(x)$ is the zeroth order term
in $\Omega$, while the terms containing the functions $k_1 (x)$ and $k_2 (x)$
are the 1st order terms in $\Omega$.
The $D_{ab}$'s as functions of the collective coordinates $\xi_A$ are the
standard Wigner rotation matrices, while $R_a$ is the right rotation
generator familiar in the SU(3) Skyrme model. In the above expressions, 
${\langle O \rangle}_B$ should be understood
as an abbreviated notation of the matrix element of a collective
operator $O$ between a baryon state $B$ with appropriate quantum
numbers, i.e.
\begin{equation}
 {\langle O \rangle}_B \equiv \int
 \Psi_{Y T T_3 ; J J_3}^{(n)*} [\xi_A] \,O [\xi_A] \,
 \Psi_{Y T T_3 ; J J_3}^{(n)} [\xi_A] \,d \xi_A .
\end{equation}
The relevant matrix elements of the collective space operators between
the nucleon state, appearing in the above expressions, are already given
in Eqs.(186)-(188) of the paper \cite{Wakamatsu03B}. 

There are two types of CSV corrections to the distributions, $q^{(0)}(x)$,
$q^{(3)}(x)$, and $q^{(8)}(x)$. We can show that the first corrections, which
were called the dynamical plus kinematical corrections
in \cite{Wakamatsu03B} (see also \cite{BPG96},\cite{WK96}),
are given by
\begin{eqnarray}
 q^{(0)} (x \, ; \Delta m^{dyn + kin} ) \ &=& \  
 - \frac{2 \,\Delta m \,I_1}{\sqrt{3}} \,
 \left\langle D_{38} \right\rangle_N \cdot \tilde{k}_0 (x), \\
 q^{(3)} (x \ ; \Delta m^{dyn + kin} ) \ &=& \  
 - \,\frac{2 \,\Delta m_s \,I_1}{3} \,
 \left\langle D_{38} \,D_{38} \right\rangle_N \cdot
 \tilde{k}_0 (x) \nonumber \\
 &\,& \ - \,\Delta m \,I_1 \,\,
 \left\langle \sum_{i=1}^3 \,\{ D_{38}, D_{38} \} \right\rangle_N \cdot
 \left[ \tilde{k}_1 (x) - \frac{K_1}{I_1} \,k_1 (x) \right] \nonumber \\
 &\,& \ - \,\Delta \,m \,I_2 \,\,
 \left\langle \sum_{i=4}^7 \{ D_{3K}, D_{3K} \} \right\rangle_N \cdot
 \left[ \tilde{k}_2 (x) - \frac{K_2}{I_2} \,k_2 (x) \right] , \\
 q^{(8)} (x \, ; \Delta m^{dyn + kin} ) \ &=& \  
 - \,\frac{2 \Delta m \,I_1}{3} \,
 \left\langle D_{88} \,D_{88} \right\rangle_N \cdot
 \tilde{k}_0 (x) \nonumber \\
 &\,& \ - \,\Delta m \,I_1 \,\,
 \left\langle \sum_{i=1}^3 \{ D_{8i}, D_{3i} \} \right\rangle_N \cdot
 \left[ \tilde{k}_1 (x) - \frac{K_1}{I_1} \,k_1 (x) \right] \nonumber \\
 &\,& \ - \,\Delta m \,I_2 \,\,
 \left\langle \sum_{i=4}^7 \{ D_{8K}, D_{3K} \} \right\rangle_N \cdot
 \left[ \tilde{k}_2 (x) - \frac{K_2}{I_2} \,k_2 (x) \right] .
\end{eqnarray}
Here, $I_1$, $I_2$, $K_1$, and $K_2$ are various moments of inertia of
the soliton defined through Eqs.(49)-(52) in the paper \cite{Wakamatsu03B}.
On the other hand, the functions $\tilde{k}_0 (x)$, $\tilde{k}_1 (x)$
and $\tilde{k}_2 (x)$ are respectively given in Eqs.(142), (155), and (156)
in the same paper \cite{Wakamatsu03B}.

The necessary matrix elements of the collective space operators between
the proton state can easily be calculated and they are shown in
Table \ref{Table:Matrix_elements}. 
The three matrix elements in the left column take the same values also
for the neutron state, whereas the four matrix elements
in the right column changes signs for the neutron state.

\vspace{3mm}
\begin{table}[h]
\caption{The matrix elements of the relevant collective space operators
in the proton state.}
\label{Table:Matrix_elements}
\vspace{2mm}
\begin{center}
\renewcommand{\arraystretch}{1.0}
\begin{tabular}{c|c}
\hline\hline
 --- & 
 $\left\langle \frac{D_{38}}{\sqrt{3}} \right\rangle_p \ = \ \frac{1}{30}$ \\
 \hline
 $\left\langle D_{38} \,D_{38} \right\rangle_p \ = \ \frac{1}{15}$ &
 $\left\langle D_{88} \,D_{38} \right\rangle_p \ = \ 0$ \\ \hline
 $\left\langle \sum_{i=1}^3 \,\left\{ D_{3i}, D_{3i} \right\} \right\rangle_p
 \ = \ \frac{10}{9}$ &
 $\left\langle \sum_{i=1}^3 \,\left\{ D_{8i}, D_{3i} \right\} \right\rangle_p
 \ = \ \frac{2 \,\sqrt{3}}{45}$ \\ \hline
 \ \ \ $\left\langle \sum_{K=4}^7 \,\left\{ D_{3K}, D_{3K} \right\} \right\rangle_p
 \ = \ \frac{34}{45}$ \ \ \ &
 \ \ \ $\left\langle \sum_{K=1}^3 \,\left\{ D_{8K}, D_{3K} \right\} \right\rangle_p
 \ = \ - \,\frac{2 \,\sqrt{3}}{45}$ \ \ \ \\ 
 \hline \hline
\end{tabular}
\end{center}
\end{table}

The second correction to the PDFs arises from the mixing of the
SU(3) representation by the CSV mass
term \cite{Wakamatsu03B},\cite{BPG96},\cite{WK96}.
Due to the presence of the CSV
mass term, the nucleon state is not a pure SU(3) octet, but it is a
linear combination of three SU(3) representation as
\begin{equation}
 | \,N \rangle \ \simeq \ |\,8, N \rangle \ + \ 
 d^N_{\bar{10}} \,|\,\bar{10}, N \rangle \ + \ 
 d^N_{27} \,|\,27, N \rangle ,
\end{equation}
with the mixture constants,
\begin{eqnarray}
 d^N_{\bar{10}} \ &=& \ \frac{\sqrt{5}}{15} \,\left( \alpha^\prime \ + \ 
 \frac{1}{2} \,\gamma^\prime \right) \,I_2 , \\
 d^N_{27} \ &=& \ - \,\frac{\sqrt{6}}{75} \,\left( \alpha^\prime \ - \ 
 \frac{1}{6} \,\gamma^\prime \right) \,I_2 . 
\end{eqnarray}
Here, the constants $\alpha^\prime$ and $\gamma^\prime$ are given by
\begin{eqnarray}
 \alpha^\prime &=& \left( \frac{\bar{\sigma}}{N_c} \ - \ \frac{K_2}{I_2} \right) \,
 \frac{\Delta m}{2} , \\
 \gamma^\prime &=& - \,\left( \frac{K_1}{I_1} \ - \ \frac{K_2}{I_2} \right) 
 \,\Delta ,
\end{eqnarray}
with $N_c = 3$ being the number of colors, whereas 
$\bar{\sigma}$ is defined in Eq.(206) of \cite{Wakamatsu03B}.

\begin{figure}[ht]
\begin{center}
\includegraphics[width=10.0cm]{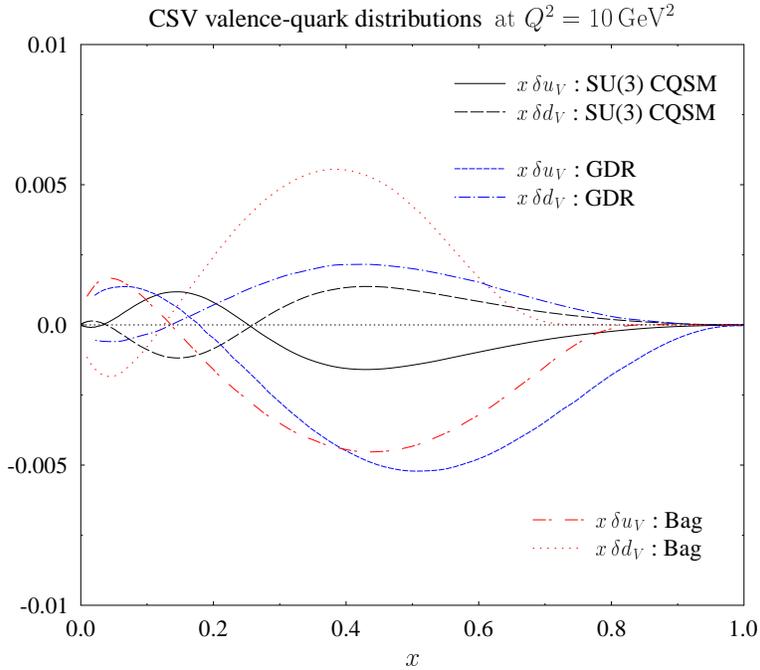}
\caption{The predictions of the SU(3) CQSM for the CSV PDFs $\delta u_V (x)$
and $\delta d_V (X)$ corresponding $Q^2 = 10 \,\mbox{GeV}^2$. Also shown
are the bag model predictions by Rodionov, Thomas, and
Londergan \cite{RTL94}, and the
predictions based on the QED radiative mechanism by
Gl\"{u}ck et al. \cite{GJR05}}
\label{Fig16_GDR_CSV_val}
\end{center} 
\end{figure}

Putting all the functions above together, the CSV corrections to the PDFs
can be evaluated in the following manner : 
\begin{eqnarray}
 \delta u(x) \ \equiv \ u^p(x) - d^n(x) &=& 
 \ \frac{2}{3} \,\,\delta q^{(0)} (x ; \Delta m^{dyn+kin}) \nonumber \\
 &+& \left[\, \delta q^{(3)} (x ; \Delta m^{dyn+kin}) \ + \ 
 \delta q^{(3)} (x ; \Delta m^{rep}) \,\right] \nonumber \\
 &+& \frac{1}{\sqrt{3}} \,\left[\, \delta q^{(8)} (x ; \Delta m^{dyn+kin})
 \ + \ \delta q^{(8)} (x; \Delta m^{rep}) \,\right] , \\
 \delta d(x) \ \equiv \ d^p(x) - u^n(x) &=& 
 \ \frac{2}{3} \,\,\delta q^{(0)} (x ; \Delta m^{dyn+kin}) \nonumber \\
 &-& \left[\, \delta q^{(3)} (x ; \Delta m^{dyn+kin}) \ + \ 
 \delta q^{(3)} (x ; \Delta m^{rep}) \,\right] \nonumber \\
 &+& \frac{1}{\sqrt{3}} \,\left[\, \delta q^{(8)} (x ; \Delta m^{dyn+kin})
 \ + \ \delta q^{(8)} (x; \Delta m^{rep}) \,\right] , \\
 \delta s(x) \ \equiv \ s^p(x) - s^n(x) &=& 
 \ \frac{2}{3} \,\,\delta q^{(0)} (x ; \Delta m^{dyn+kin}) \nonumber \\
 &-& \frac{2}{\sqrt{3}} \,\left[\, \delta q^{(8)} (x ; \Delta m^{dyn+kin})
 \ + \ \delta q^{(8)} (x; \Delta m^{rep}) \,\right] .
\end{eqnarray}
Now, we show in Fig.\ref{Fig16_GDR_CSV_val} the predictions of
the SU(3) CQSM for the CSV
PDFs evolved to the scale $Q^2 = 10 \,\mbox{GeV}^2$ in comparison
with some other theoretical predictions. 
The solid and long-dashed curves (black in color) respectively stand
for the predictions of the SU(3) CQSM for
$x \,\delta u_V \equiv x \,[u_V^p (x) - d_V^n (x)]$ and
$x \,\delta d_V \equiv x \,[d_V^p (x) - u_V^n (x)]$. 
The long dash-dotted and dotted curves (red in color)
are the predictions of Rodionov, Thomas, and Londergan based on the bag model
with quark-diquark correlations \cite{RTL94}.
On the other hand, the short-dashed
and short dash-dotted (blue in color) are the predictions of
Gl\"{u}ck, Jimenez-Delgado, and Reya based on the QED radiative (or splitting)
mechanism \cite{GJR05}. (The CSV effects arising from the QED splitting
mechanism was also proposed by Martin et al. independently \cite{MRSTCSV05}.   

\vspace{5mm}

First, we point out that all the models predicts that $\delta u_V (x) < 0$
and $\delta d_V (x) > 0$ at least for the dominant components in the
larger $x$ region. Comparing the predictions of the SU(3) CQSM and those
of the bag model, we find that the former are much smaller than the
latter. 
To understand the cause of this difference, it is
instructive to compare the basic framework of
these models in some detail. In the framework of the CQSM, the
mass difference between the $u$- and $d$-quarks is the only origin of
the CSV effects in the PDFs. Once the perturbative treatment of this
mass difference is accepted, there is no ambiguity in the theoretical
treatment. On the other hand, refined bag model treatment of Rodionov 
et al. is based on quite a different assumption on the CSV
mechanism \cite{RTL94}, which was first proposed by
Sather \cite{Sather92} and has been used in most
investigations of the CSV PDFs in the past.
This treatment is critically dependent on the quark-diquark picture
for the intermediate states in the DIS amplitudes. 
To be more concrete, their treatment starts with
the parton model expression for a quark distribution
function given as
\begin{equation}
 q(x) \ = \ M_N \,\sum_X \,\left| \langle X \,|\, \psi_+ (0) \,|\, 
 N \rangle \right|^2 \times \delta(M_N (1-x) - p^+_X) ,
\end{equation}
where $\psi_{\pm} = (1 + \gamma^0 \,\gamma^3) \,\psi / 2$, while $| \,N \rangle$
is the nucleon state, while $|\,X \rangle$ is all possible final state,
which is obtained from $|\,N \rangle$ by removing a quark or adding
an antiquark. The state $|\,X \rangle$ is thought to have the following
Fock-space expansion, $|\,X \rangle = 2 \,q$, $3 \,q + \bar{q}$,
$4 \,q + 2 \,\bar{q}, \cdots$.
Based on the idea that, for large enough $x$, say, $x \ge 0.2$, 
the valence quarks dominate, it is postulated that a reasonable
estimate of $q(x)$ can be obtained by including only two-quark
intermediate states for $|\,X \rangle$. It is further assumed that
this intermediate two-quark state can be approximated by a diquark with
definite mass $M_D$. 
The validity of both these assumptions is not absolutely clear.
Especially, the latter postulate, i.e. the two-body kinematics in the
intermediate state, is a highly nontrivial assumption.
We refer to the paper \cite{CS01} by Cao and Signal for the detailed criticism
to the framework of evaluating the CSV effects in PDFs based on the
quark-diquark hypothesis.

In any case, a common feature of most calculations based on
this quark-diquark picture is that they predict fairly large CSV
corrections in PDFs ranging from 2\% to 10\%, which is much larger than
the CSV effects expected from the low energy CSV phenomena, which
are generally known to be less than 1\%. 
In view of this situation, it is important to estimate the size of CSV
effects in PDFs without
relying upon the quark-diquark picture.
So far, there have been only a few such attempts. The one is the study
by Cao and Signal based on a meson cloud model \cite{CS01}.
In their treatment,
hadron mass differences between the isospin multiplets are the 
only sources of the CSV effects in PDFs.
Another independent analysis was carried out by Benesh and Goldman based on
a quark model \cite{BG97}.
In their treatment, the effects due to the $u$-$d$ quark
mass difference and the Coulomb interaction of the electrically charged
quarks are taken into account. Both of these studies shows that the
CSV effects in PDFs are considerably smaller than those obtained based on the
quark-diquark picture. The present calculation based on a totally
different theoretical framework appears to give another support to
this conclusion by Cao and Signal and also by Benesh and Goldman.
 
\begin{figure}[ht]
\begin{center}
\includegraphics[width=9.0cm]{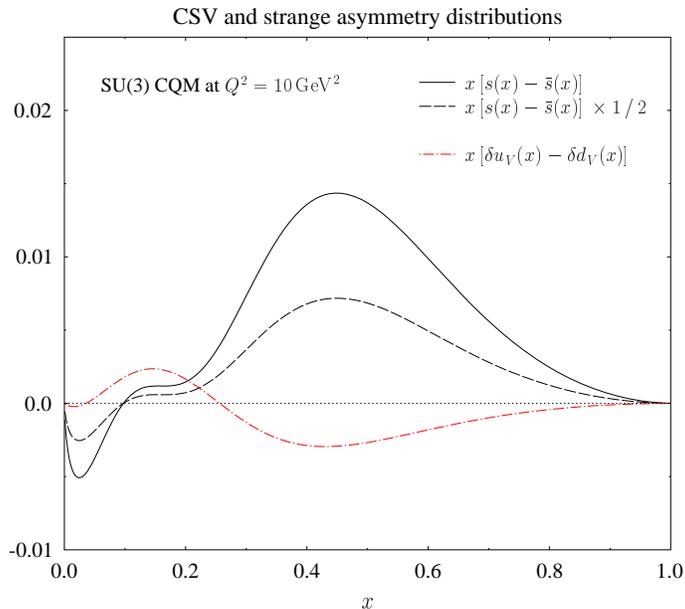}
\caption{The SU(3) CQSM predictions for the CSV valence quark distribution
in comparison with the strange asymmetry distribution.}
\label{Fig17_CSV_Sasym_dist}
\end{center} 
\end{figure}

Since our main purpose of investigating the CSV PDFs is to get a
feeling about the relative importance the CSV effects and the
asymmetry of the strange and antistrange quark distributions 
in the resolution scenario of the NuTeV anomaly, we compare
these distributions in Fig.\ref{Fig17_CSV_Sasym_dist}.
Here, the solid curve is the bare
prediction of the CQSM for $x \,[s(x) - \bar{s}(x)]$, while the
long-dashed curve is the reduced prediction by a
factor of $1/2$. (The latter is our favorable prediction, as explained
before.) The dash-dotted curve (red in color) is the CQSM prediction
for the CSV valence distribution $x \,[\delta u_V (x) - \delta d_V (x)]$
divided by a factor of $2$. (Compare Eq.(\ref{Eq:RI}) and Eq.(\ref{Eq:RS})
below for the reason why we divide it by $2$.)
One sees that the CSV valence quark
distribution is much smaller than the asymmetry of strange and antistrange
quark distribution calculated within exactly the same theoretical
framework.

\vspace{5mm}

As is well-known, the main QCD correction to the Paschos-Wolfenstein ration
is approximately given by the following formula : 
\begin{equation}
 R^- \ \equiv \ \frac{\sigma^{\nu N}_{NC} - \sigma^{\bar{\nu} N}_{NC}}
 {\sigma^{\nu N}_{CC} - \sigma^{\bar{\nu} N}_{CC}} \ = \ 
 R^-_{PW} \ + \ \delta R^-_I \ + \ \delta R^-_s ,
\end{equation}
where
\begin{eqnarray}
 \delta R^-_I \ &\simeq& \ \left( 1  \ - \ \frac{7}{3} \,s^2_W \right) \,
 \frac{\delta U_V - \delta D_V}{2 \,(U_V + D_V)} , \label{Eq:RI} \\
 \delta R^-_s \ &\simeq& \ - \,\left( 1 \ - \ \frac{7}{3} \,s^2_W \right) \,
 \frac{S^-}{U_V + D_V}, \label{Eq:RS}
\end{eqnarray}
with $s^2_W \equiv \sin^2 \theta_W = 0.2227 \pm 0.0004$ and
\begin{eqnarray}
 Q_V (Q^2) &=& \int_0^1 \,x \,q_V (x,Q^2) \,dx \\
 \delta Q_V (Q^2) &=& \int_0^1 \,x \,\delta q_V (x,Q^2) \,dx, \\
 S^- (Q^2) &=& \int_0^1 \, x \,[s(x,Q^2) - \bar{s}(x,Q^2) \,] \,dx .
\end{eqnarray}
Gl\"{u}ck et al. \cite{GJR05} as well as the NuTeV
group \cite{NuTeV02A},\cite{NuTeV02B} pointed out that the above
approximate formula is
not accurate enough and proposed more refined formula to determine
the CSV effects on the determination of $\sin \theta_W$.
However, since our main interest here is the relative importance
of the CSV effects and the particle-antiparticle asymmetry of the
strange quark distribution, we use the above formula below.
Using the obtained distributions corresponding to the scale
$Q^2 = 10 \,\mbox{GeV}^2$, we get the following estimate 
for the CSV correction of QCD origin : 
\begin{equation}
 \Delta s^2_W |_{CSV} \ \simeq \ \delta R^-_I |_{QCD} \ \simeq \ 
 - \,0.00035 .
\end{equation}
On the other hand, the correction due to the strange-antistrange
asymmetry is given by
\begin{equation}
 \Delta s^2_W |_{strange} \ \simeq \ \delta R^-_s \ = \ - \,0.00264 \,\,
 (- \,0.00528) .
\end{equation}
Here, the number in the parenthesis is the bare prediction of the
SU(3) CQSM not being multiplied by a reduced factor of $1/2$.
Note that this estimate is order of magnitude consistent with
the independent estimate by Ding, Xu, and Ma \cite{DXM05},
which gives $\delta R^-_s \simeq - \,(0.00297 - 0.00498)$.  
One confirms that the effect of CSV originating from the $u$-$d$ quark
mass difference is order of magnitude smaller than that of the
strange asymmetry. 
We however recall that there is another mechanism
which generate the CSV effects in the quark distributions.
It is the QED splitting mechanism proposed by Gl\"{u}ck et al. and
also by Martin et al. The recent estimate by Gl\"{u}ck et al.
gives
\begin{equation}
 \Delta s^2_W |_{QED} \ = \ \delta R^-_I |_{QED} \ = \ - \,0.002.
\end{equation}
Since this CSV mechanism is of QED origin and it is totally independent of
the CSV effect of QCD origin, we may add all the above corrections
to the Weinberg angle. This gives
\begin{eqnarray}
 \Delta s^2_W |_{sum} &=& \mbox{QED} \ + \ \mbox{Strange} \ + \ 
 \mbox{CSV} \nonumber \\
 &=& - \,0.002 \ - \ 0.00264 \ - \ 0.00035 \nonumber \\
 &\simeq& - \,0.0050 .
\end{eqnarray}
This means that the NuTeV measurement of $\sin^2 \theta_W = 0.2277 \,(16)$
will be shifted to $\sin^2 \theta_W = 0.2227 \,(16)$ which agree with the
standard value $0.2228 \,(4)$, although we should perform more careful
analysis in view of the approximate nature of the above correction formula. 
Anyhow, our finding here can be summarized as follows.
The effect of the particle-antiparticle asymmetry of the strange quark
distribution to the NuTeV anomaly seems to be much larger than the CSV
effect in the valence quark distribution originating from
the $u$-$d$ quark mass difference. However, the CSV effect due to the
QED splitting mechanism is an increasing function of the
scale \cite{GJR05},\cite{MRSTCSV05} and
its effect on the NuTeV anomaly can have the same order of magnitude
as that of the strange asymmetry at the scale of $Q^2 = 10 \,\mbox{GeV}^2$.

\vspace{5mm}
\begin{figure}[ht]
\begin{center}
\includegraphics[width=9.0cm]{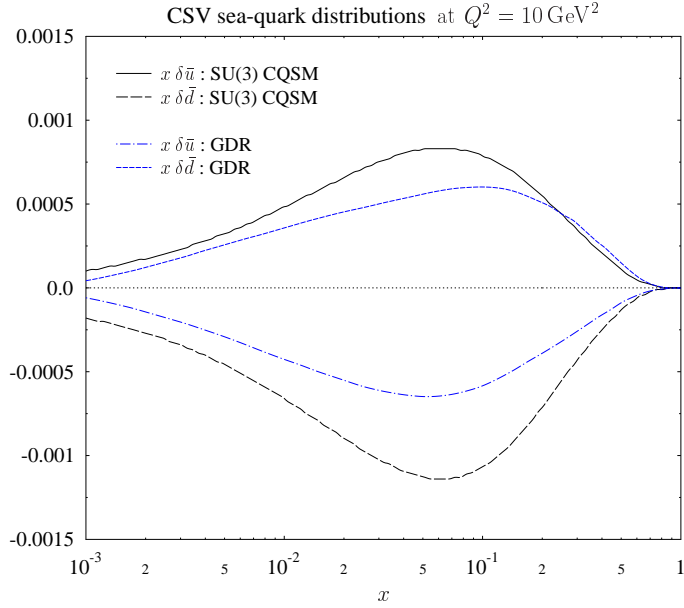}
\caption{The predictions of the SU(3) CQSM for the CSV sea-quark
(antiquark) distributions at $Q^2 = 10 \,\mbox{GeV}^2$ in
comparison with the corresponding distributions generated by
the QED splitting mechanism \cite{GJR05}.}
\label{Fig18_GDR_CSV_sea}
\end{center} 
\end{figure}

Since one of the advantages of the CQSM is that it can give
reasonable predictions not only for the quark distributions but also
for the antiquark distributions, we think it interesting to evaluate
the CSV effect in the sea-quark distributions in this model.
The solid and long-dashed curves in
Fig.\ref{Fig18_GDR_CSV_sea} represent
the CSV light-flavor sea-quark distributions defined by
\begin{eqnarray}
 \delta \bar{u}(x) \ &\equiv& \ \bar{u}^p (x) \ - \ \bar{d}^n (x), \\
 \delta \bar{d}(x) \ &\equiv& \ \bar{d}^p (x) \ - \ \bar{u}^n (x).
\end{eqnarray}
Here, the solid and long-dashed curves (black in color) correspond to
the predictions of the SU(3) CQSM, while the dash-dotted and short-dashed
curves (blue in color) correspond to the predictions based on
the QED splitting mechanism \cite{GJR05}.   
Very curiously, the predictions of the SU(3) CQSM for $\delta \bar{u}(x)$
and $\delta \bar{d}(x)$ and the corresponding predictions due to the QED splitting
mechanism have nearly equal magnitudes but their signs are opposite.
This means that, if we add up both contributions, a sizable cancellation
occurs, which would indicate that the net CSV effects
on the sea-quark distribution would be very small and hard to
observe experimentally.

\vspace{5mm}
\begin{figure}[ht]
\begin{center}
\includegraphics[width=9.0cm]{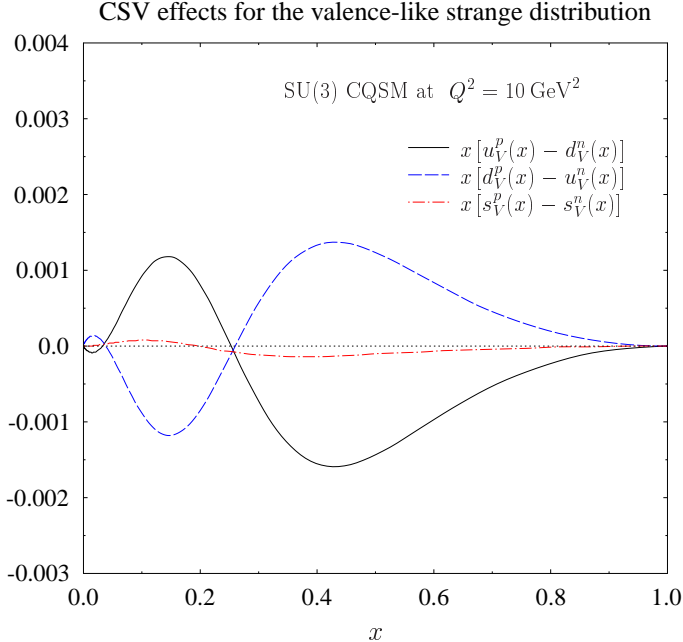}
\caption{The SU(3) CQSM prediction for the CSV effects to the
valence-like strange quark distribution $x \,[s^p_V (x) - s^n_V (x)]$
with $s_V(x) \equiv s(x) - \bar{s}(x)$ in comparison with the
CSV effects to the light-flavor valence-like distributions
$x \,[u^p_V (x) - d^n_V (x)]$ and $x \,[d^p_V (x) - u^n_V (x)]$.}
\label{Fig19_CSV_uds_val}
\end{center} 
\end{figure}

Finally, just to make sure, we estimate the CSV effect on the valence-like
strange quark distribution, i.e. $s^- (x) \equiv s(x) - \bar{s}(x)$
in comparison with the CSV effects on the light-flavor valence
quark distribution. The results are shown in Fig.\ref{Fig19_CSV_uds_val}.
One simply confirms that the CSV
effect on the strange distribution is in fact very small. 

\section{Summary and Conclusion}

To conclude, we have analyzed the unpolarized and the longitudinally polarized
PDFs in the nucleon within a single theoretical framework of the
SU(3) CQSM, which contains only one adjustable parameter $\Delta m_s$,
the mass difference between the strange and nonstrange quarks.
Through detailed comparisons with the recent global PDF fits by
the NNPDF group, the DSSV group, and the CTEQ group, etc., we could
confirm that, despite its nearly parameter-free nature, the model
reproduces all the qualitative characteristics of the empirically
determined PDFs. Besides, it gives unique and nontrivial predictions on
the flavor structure of the sea-quark distributions.
They are the flavor asymmetry of the
unpolarized sea-quark distributions, $\bar{u}(x) - \bar{d}(x) < 0$,
dictated by the famous NMC measurement, the flavor asymmetry of
the longitudinally polarized sea-quark distributions,
$\Delta \bar{u}(x) > 0$, $\Delta \bar{d}(x) < 0$, the
particle-antiparticle asymmetry of the unpolarized strange quark distribution,
$s(x) - \bar{s}(x) \neq 0$, and also the particle-antiparticle asymmetry
of the longitudinally polarized strange quark distributions,
$\Delta s(x) < 0$, $\Delta \bar{s}(x) \simeq 0$. 
The success is naturally connected with the fact that the model incorporates
the most important feature of the QCD in the nonperturbative low energy domain,
i.e. the spontaneous chiral symmetry breaking and
the appearance of the associated Goldstone bosons.
Still, an important difference with more familiar meson cloud
models should be clearly recognized. As stated above,
the CQSM predicts large isospin asymmetry not only for the
unpolarized seas but also for the longitudinally polarized ones.
On the other hand, although the meson cloud models nicely explains the flavor
asymmetry of the unpolarized sea-quark distributions, they generally predict
very small spin polarization of sea quarks, reflecting the fact that the pion
carries no spin and the effects of heavier meson cloud are suppressed.
In view of this important difference, more unambiguous 
confirmation of the flavor asymmetry of the longitudinally polarized sea-quark
distribution is an urgent task. 

We have pointed out that, for that the model predictions are taken
as reliable also in a quantitative sense, we need two remedies.
First, the information from phenomenological global fits indicates
that the gluon carries about 20 \% of the nucleon momentum fraction even
at the low energy scale corresponding to the CQSM.
Naturally, this fact is not properly incorporated in effective quark models like the
CQSM. As we have seen, this seems to be a cause of about 20 \% overestimate
of the flavor-singlet combination of the unpolarized PDFs.
However, we have also shown that the neglect of the gluon degrees of freedom
at the model energy scale is likely to do little harm in the case of the longitudinally polarized PDFs. This is the reason why the success of
the model is more salient for the longitudinally polarized PDFs than
for the unpolarized PDFs. The second problem is that the SU(3) symmetric
collective quantization (with the subsequent perturbative treatment of
the SU(3) symmetry breaking mass difference between the
strange and nonstrange quarks) might tend to overestimate the kaon cloud effects,
thereby having a danger of overvaluing the magnitudes of the strange
quark distributions. As the present analysis, especially the detailed
comparison with the unbiased NNPDF global fits, strongly indicates,
plausible predictions for the strange and antistrange quark distributions
would corresponds to an average of the SU(3) and SU(2) CQSM, which means that
we can get reliable predictions for the strangeness-related distributions
if we multiply a reduction factor of about $1/2$ to the bare predictions
of the SU(3) CQSM.
After this modification to the strange quark distributions is taken into account,
we have good reason to believe that the SU(3) CQSM is already giving reliable
predictions with $(20-30) \%$ accuracy for both of the unpolarized and
longitudinally polarized PDFs, including the key issues of the present
research, i.e. the flavor structure of the sea-quark distribution in the nucleon. We hope that these characteristic predictions reported in the present paper
will be tested through more elaborate analyses of the
neutrino-induced DIS measurements, the semi-inclusive DIS measurements,
the polarized Drell-Yan processes in $pp$ or $p \bar{p}$ collisions, etc.,
to be carried out in the near future.

\vfill

\vspace{-3mm}

\begin{acknowledgments}
The author would like to thank Y.~Nakakoji for his helpful collaboration
at the early stage on the investigation of the charge-symmetry-violating
parton distributions. He also greatly acknowledges useful discussions with
J.-C.~Peng and K.-F.~Liu at the workshop ``Flavor Structure of the Nucleon Sea''
held at ECT* in July, 2013.
\end{acknowledgments}

\vspace{3mm}

\begin{thebibliography}{10}

%











\bibitem{Muta98}
T.~Muta, {\it Foundations of Quantum Chromodynamics}
(World Scientific, Singapore, 1998).

\bibitem{Collins11}
J.~Collins, {\it Foundations of Perturbative QCD}
(Cambridge University Press, New York, 2011).

\bibitem{AEL95}
M.~Anselmino, A.~Efremov, and E.~Leader, 
Phys. Rep. {\bf 261}, 1 (1995).

\bibitem{LR00}
B.~Lampe and E.~Reya, Phys. Rep. {\bf 332}, 1 (2000).

\bibitem{JMO13}
P.~Jimenez-Delgado, W.~Melnitchouk, and F.~Owens,
J. Phys. G : Nucl. Part. Phys. {\bf 40}, 093102 (2013).



\bibitem{NMC91}
NMC Collaboration, P.~Amaudruz et al., 
Phys. Rev. Lett. {\bf 66}, 2712 (1991).

\bibitem{Kumano98}
S.~Kumano, Phys. Rep. {\bf 303}, 183 (1998).

\bibitem{GP01}
G.T.~Garvey and J.-C. Peng,
Prog. Part. Nucl. Phys. {\bf 47}, 204 (2001).

\bibitem{PQ14}
J.-C.~Peng and J.-W.~Qiu,
Prog. Part. Nucl. Phys. {\bf 76}, 43 (2014).


\bibitem{Sullivan72}
J.D.~Sullivan, Phys. Rev. D {\bf 5}, 1732 (1972).

\bibitem{Aubert83}
J.J.~Aubert et al., Phys. Lett. B {\bf 123}, 275 (1983).

\bibitem{Kumano91}
S.~Kumano, Phys. Rev. D {\bf 43}. 3067 (1991).

\bibitem{KL91}
S.~Kumano and J.T.~Londergan, Phys. Rev. D {\bf 44}, 717 (1991).

\bibitem{Wakamatsu91}
M.~Wakamatsu, Phys. Rev. D {\bf 44}, R2631 (1991).

\bibitem{Wakamatsu92}
M.~Wakamatsu, Phys. Rev. D {\bf 46}, 3762 (1992).

\bibitem{HSB91}
W.-Y.P.~Hwang, J.~Speth, and G.E.~Brown, Z. Phys. A {\bf 339}, 383 (1991).

\bibitem{KFS96}
W.~Koepf, L.L.~Frankfurt, M.~Strikman, Phys. Rev. D {\bf 53}, 2586 (1996).



\bibitem{CCFR95}
CCFR Collabotation, A.~Bazarko et al,
Z. Phys. C {\bf 65}, 189 (1995).

\bibitem{CCFRNuTeV01}
CCFR/NuTeV Collaboration, Un-Ki~Yang et al., Phys. Rev. Lett. {\bf 86},
2742 (2001).

\bibitem{BPZ00}
V.~Barone, C.~Pascaud, and F.~Zomer,
Eur. Phys. J. C {\bf 12}, 243 (2000).

\bibitem{NNPDF09}
NNPDF Collabotation, R.D.~Ball et al.,
Nucl. Phys. B {\bf 823}, 195 (2009).

\bibitem{AKP09}
A.~Alekhin, S.~Kulagin, and R.~Petti, 
Phys. Lett. B {\bf 675}, 433 (2009).

\bibitem{SYKMOO08}
I.~Schienbein, J.Y.~Yu, C.~Keppel, J.G. Morf\'in, F.I.~Olness,
and J.F.~Owens, \\
Phys. Rev. D {\bf 77}, 054013 (2008).

\bibitem{SYKKMOO09}
I.~Schienbein, J.Y.~Yu, K.~Kova\v{r}\'ik, C.~Keppel, J.G. Morf\'in,
F.I.~Olness, and J.F.~Owens, \\
Phys. Rev. D {\bf 80}, 094004 (2009).


\bibitem{HERMES08}
HERMES Collaboration, A.~Airapetian et al.,
Phys. Lett. B {\bf 666}, 446 (2008).

\bibitem{HERMES05}
HERMES Collaboration, A.~Airapetian et al., 
Phys. Rev. D {\bf 71}, 012003 (2005).

\bibitem{HERMES99}
HERMES Collaboration, K.~Ackerstaff et at.,
Phys. Lett. B {\bf 464}, 123 (1999).

\bibitem{COMPASS09}
COMPASS Collaboration, M.~Alekseev et al., 
Phys. Lett. B {\bf 680}, 217 (2009).

\bibitem{COMPASS08}
COMPASS Collaboration, M.~Alekseev et al.,
Phys. Lett. B {\bf 660}, 458 (2008).


\bibitem{Kretzer00}
D. de~Florian, R.~Sassot, M.~Stratmann, Phys. Rev. D {\bf 75},
114010 (2007).

\bibitem{DSSV07A}
D. de~Florian, R.~Sassot, M.~Stratmann, Phys. Rev. D {\bf 76},
074033 (2007).

\bibitem{DSSV07B}
S.~Kretzer, Phys. Rev. D {\bf 62}, 054001 (2000).

\bibitem{HKNS07}
M.~Hirai, S.~Kumano, T.-H. Nagai, and K.~Sudoh, Phys. Rev.
D {\bf 75}, 094009 (2007).

\bibitem{AKK08}
S.~Albino, B.A.~Kniehl, and G.~Krammer, Nucl Phys. B {\bf 803}, 42 (2008).

\bibitem{Belle08}
Belle Collaboration, R.~Seidl et al., Phys. Rev. D {\bf 78}, 032011 (2008).


\bibitem{DPP88}
D.I.~Diakonov, V.Yu.~Petrov, and P.V.~Pobylitsa, Nucl. Phys. B {\bf 306},
809 (1988).

\bibitem{WY91}
M.~Wakamatsu and H.~Yoshiki, Nucl. Phys. A {\bf 524}, 561 (1991).

\bibitem{KR84}
S.~Kahana and G.~Ripka, Nucl. Phys. A {\bf 429}, 462 (1984).



\bibitem{Review_Wakamatsu92}
M.~Wakamatsu, Prog. Theor. Phys. Supple. {\bf 109}, 115 (1992).

\bibitem{Review_CBKPWMAG96}
Chr.V.~Christov, A.~Blotz, H.-C.~Kim, P.V.~Pobylitsa, T.~Wakabe,
Th.~Meissner, E.Ruiz~Arriola, and K.~Goeke, Prog. Theor. Nucl. Part. Phys.
{\bf 37}, 91 (1996).

\bibitem{Review_ARW96}
R.~Alkofer, H.~Reinhardt, and H.~Weigel, Phys. Rep. {\bf 265}, 139 (1996).

\bibitem{Review_DP01}
D.I.~Diakonov and V.Yu.~Petrov, {\it At the Frontier of Particle Physics}
(World Scientific, Singapore, 2001), Vol.1.


\bibitem{DPPPW96}
D.I.~Diakonov, V.Yu.~Petrov, P.V.~Pobylitsa, M.V.~Polyakov, and C.~Weiss,\\
Nucl. Phys. B {\bf 480}, 341 (1996).

\bibitem{DPPPW97}
D.I.~Diakonov, V.Yu.~Petrov, P.V.~Pobylitsa, M.V.~Polyakov, and C.~Weiss,\\
Phys. Rev. D {\bf 56}, 4069 (1997). 

\bibitem{PPGWW99}
P.V.~Pobylitsa, M.V.~Polyakov, K.~Goeke, T.~Watabe, and C.~Weiss,\\
Phys. Rev. D {\bf 59}, 034024 (1999).

\bibitem{WGR96}
H.~Weigel, L.~Gamberg, and H.Reinhardt, Mod. Phys. Lett. A {\bf 11}, 3021 (1996).

\bibitem{WGR97}
H.~Weigel, L.~Gamberg, and H.Reinhardt, Phys. Lett. B {\bf 399}, 286 (1997).

\bibitem{GRW98}
L.~Gamberg, H.~Reinhardt, and H.~Weigel, Phys. Rev. D {\bf 58}, 054014 (1998).

\bibitem{WK98}
M.~Wakamatsu and T.~Kubota, Phys. Rev. D {\bf 57}, 5755 (1998).

\bibitem{WK99}
M.~Wakamatsu and T.~Kubota, Phys. Rev. D {\bf 60}, 034020 (1999).

\bibitem{WW00}
M.~Wakamatsu and T.~Watabe, Phys. Rev. D {\bf 62}, 054009 (2000).

\bibitem{Wakamatsu03A}
M.~Wakamatsu, Phys. Rev. D {\bf 67}, 034005 (2003).

\bibitem{Wakamatsu03B}
M.~Wakamatsu, Phys. Rev. D {\bf 67}, 034006 (2003).


\bibitem{CCS10}
H.~Chen, F.-G.~Cao, and A.I.~Signal, 
J. Phys. G : Nucl. Part. Phys. {\bf 37}, 105006 (2010).

\bibitem{CS03}
F.-G.~Cao and A.I.~Signal, Phys. Lett. B {\bf 559}, 220 (2003).



\bibitem{NuTeV02A}
NuTeV Collaboration, G.P.~Zeller et al., Phys. Rev. Lett. {\bf 88}, 091802 (2002).

\bibitem{NuTeV02B}
NuTeV Collaboration, G.P.~Zeller et al., Phys. Rev. D {\bf 65}, 111103 (2002).


\bibitem{ST87}
A.I.~Signal and A.W.~Thomas, Phys. Lett. B {\bf 191}, 205 (1987).

\bibitem{BW92}
M.~Burkardt and B.J.~Warr, Phys. Rev D {\bf 45}, 958 (1992).

\bibitem{BM96}
S.J.~Brodsky and B.-Q.~Ma, Phys. Lett. B {\bf 381}, 317 (1996).

\bibitem{DM04}
Y.~Ding and B.-Q.~Ma, Phys. Lett. B {\bf 590}, 216 (2004).

\bibitem{AI04}
J.~Alwall and G.~Ingelman, Phys. Rev. D {\bf 70}, 111505 (2004).

\bibitem{Wakamatsu05}
M.~Wakamatsu. Phys. Rev. D {\bf 71}, 057504 (2005).

\bibitem{DXM05}
Y.~Ding, R.-G.~Xu, and B.-Q.~Ma,
Phys. Lett. B {\bf 607}, 101 (2005).



\bibitem{LT05}
J.T.~Londergan and A.W.~Thomas, J. Phys. G : Nucl. Part. Phys. {\bf 31},
1151 (2005).

\bibitem{LPT10}
J.T.~Londergan, J.C.~Peng, and A.W.~Thomas, Rev. Mod. Phys. {\bf 82},
2009 (2010).


\bibitem{Sather92}
Eric Sather, Phys. Lett. B {\bf 274}, 433 (1992).

\bibitem{RTL94}
E.~Rodionov, A.W.~Thomas, and J.T.~Londergan, Mod. Phys. Lett. A {\bf 9},
1799 (1994).

\bibitem{LT03}
J.T.~Londergan and A.W.~Thomas, Phys. Lett. B {\bf 558}, 132 (2003).


\bibitem{GJR05}
M.~Gl\'{u}ck, P.~Jimenez-Delgado, and E.~Reya, Phys. Rev. Lett. {\bf 95},
022002 (2005).

\bibitem{MRSTCSV05}
A.D.~Martin, R.G.~Roberts, W.J.~Stirling, and R.S.~Thorne,
Eur. Phys. J. C {\bf 39}, 155 (2005).


\bibitem{CSSM11}
CSSM and QCDSF/UKQCD Collaborations, R.~Horsley et al.,\\
Phys. Rev. D {\bf 83}, 051501(R) (2011).


\bibitem{NNPDFNLO2.1}
NNPDF Collaboration, R.D.~Ball et al., Nucl. Phys. B {\bf 855}, 153 (2012).


\bibitem{NNPDFpol1.0}
NNPDF Collaboration, R.D.~Ball et al, Nucl. Phys. B {\bf 855}, 153 (2012).




\bibitem{BDGPPP93}
A.~Blotz, D.~Diakonov, K.~Goeke, N.W.~Park, V.~Petrov, and P.V.~Pobylitsa,\\
Nucl. Phys. A {\bf 555}, 765 (1993).




\bibitem{Witten83}
E.~Witten, Nucl. Phys. B {\bf 223}, 422, 433 (1983).

\bibitem{NMP84}
M.A.~Nowak, P.O.~Mazur, and M.~Praszalowicz, Phys. Lett. B {\bf 147},
137 (1984).

\bibitem{Guadagnini85}
E.~Guadagnini, Nucl. Phys. B {\bf 236}, 35 (1985).


\bibitem{DP86}
D.I.~Diakonov and V.Yu.~Petrov, Nucl. Phys. B {\bf 272}, 457 (1986).

\bibitem{Diakonov03}
D.I.~Diakonov, Prog. Part. Nucl. Phys. {\bf 51}, 173 (2003).

\bibitem{PPPBGW98}
V.Yu.~Petrov, P.V.~Pobylitsa, M.V.~Polyakov, I.~B\"{o}ring, K.~Goeke,
and C.~Weiss, \\
Phys. Rev. D {\bf 57}, 4325 (1998).


\bibitem{MRST04}
A.D.~Martin, R.G.~Roberts, W.J.~Stirling, and R.S.~Thorne,
Eur. Phys. J. C {\bf 35}, 325 (2004).

\bibitem{MST06}
A.D.~Martin, W.J.~Stirling, and R.S.~Thorne,
Phys. Lett. B {\bf 636}, 259, (2006).


\bibitem{MP95}
P.J.~Mulders and S.J.~Pollock, Nucl. Phys. A {\bf 588}, 876 (1995).

\bibitem{Arriola98}
E.~Ruiz Arriola, Nucl. Phys. A {\bf 641}, 461 (1998).





\bibitem{YA88}
H.~Yabu and K.~Ando, Nucl. Phys. B {\bf 301}, 601 (1988).

\bibitem{CHK88}
C.G.~Callen, K.~Hornbostel, and I.~Klebanov, Phys. Lett. B {\bf 202}, 269 (1988).

\bibitem{BRS88}
J.P.~Blaizot, M.~Rho, and N.N.~Sccocola, Phys. Lett. B {\bf 209}, 27 (1988).



\bibitem{CP11A}
W.-C.~Chang and J.-C.~Peng, Phys. Lett. B {\bf 666}, 446 (2011).

\bibitem{LCCP12}
K.-F.~Liu, W.-C.~Chang, H.-Y.~Cheng, and J.-C.~Peng,
Phys. Rev. Lett. {\bf 109}, 252002 (2012).

\bibitem{BHPS80}
S.J.~Brodsky, P.~Hoyer, C.~Peterson, and N.~Sakai,
Phys. Lett. B {\bf 93}, 451 (1980).

\bibitem{BPS81}
S.J.~Brodsky, C.~Peterson, and N.~Sakai, Phys. Rev. D {\bf 23},
2745 (1981).


\bibitem{CTEQ08}
CTEQ Collaboration, P.M.~Nadolsky et al.,
Phys. Rev. D {\bf 78}, 013004 (2008).

\bibitem{CT10}
H.L.~Lai, M.~Guzzi, J.~Huston, Z.~Li, P.M.~Nadolsky, J.~Pumplin, and
C.P.~Yuan, Phys. Rev. D82, 074024 (2010). 

\bibitem{LSS14}
E.~Leader, A.V.~Sidorov, and D.B.~Stamenov, arXiv : 1406.4678 [hep-ph] (2014).

\bibitem{HERMES14}
HERMES Collaboration : A.!Airapetian et al., Phys. Rev. D {\bf 89}, 097101
(2014).

\bibitem{Stolarski14}
M.~Stolarski, arXiv : 1407.3721 [hep-ph] (2014).

\bibitem{CP11B}
W.-C.~Chang and J.-C.~Peng, Phys. Rev. Lett. {\bf 106}, 252002 (2011).

\bibitem{LD94}
K.-F.~Liu and S.J.~Dong, Phys. Rev. Lett. {\bf 72}, 1790 (1994).

\bibitem{Liu00}
K.-F.~Liu, Phys. Rev. D {\bf 62}, 074501 (2000).



\bibitem{EGP00}
A.V.~Efremov, K.~Goeke, and P.V.~Pobylitsa, Phys. Lett. B {\bf 488},
182 (2000).


\bibitem{DSSV14}
D.~de Florian, R.~Sassot, M.~Stratmann, and W.~Vogelsang,
arXiv:1404.4293 [hep-ph] (2014).

\bibitem{DSSV08}
D. de~Florian, R.~Sassot, M.~Stratmann, and W.~Vogelsang,\\
Phys. Rev. Lett. {\bf 101}, 072001 (2008).

\bibitem{DSSV09}
D. de~Florian, R.~Sassot, M.~Stratmann, and W.~Vogelsang,
Phys. Rev. D {\bf 80}, 034030 (2009).



\bibitem{DGPW00}
B.~Dressler, K.~Goeke, M.V.~Polyakov, and C.~Weiss,
Eur. Phys. J. C {\bf 14}, 147 (2000)


\bibitem{BPG96}
A.~Blotz, M.~Praszalowicz, and K.~Goeke,
Phys. Rev. D {\bf 53}, 485 (1996).

\bibitem{WK96}
M.~Wakamatsu and N.~Kaya, Prog. Theor. Phys. {\bf 95}, 767 (1996).



\bibitem{CS01}
F.-G.~Cao and A.I.~Signal, Eur. Phys. J. C {\bf 21}, 105 (2001).

\bibitem{BG97}
C.J.~Benesh and T.~Goldman, Phys. Rev. C {\bf 55}, 441 (1997).




\end{thebibliography}

\end{document}